\documentclass[structabstract]{aa}
\usepackage{amsmath,longtable,lscape,txfonts,natbib}
\usepackage[dvips]{epsfig}
\usepackage[dvips]{graphicx}
\usepackage[dvips]{color}
\usepackage{amssymb}  
\begin{document}
\def\en{E_{\nu}}
\def\eg{E_{\gamma}}
\def\ep{E_{p}}
\def\eel{E_{e}}
\def\epb{\epsilon_{p}^{b}}
\def\enb{\epsilon_{\nu}^{b}}
\def\enbG{\epsilon_{\nu,GeV}^{b}}
\def\enbM{\epsilon_{\nu,MeV}^{b}}
\def\ens{\epsilon_{\nu}^{s}}
\def\ensG{\epsilon_{\nu,GeV}^{s}}
\def\egb{\epsilon_{\gamma}^{b}}
\def\egbM{\epsilon_{\gamma,MeV}^{b}}
\def\lumi{L_{\gamma}^{52}}
\def\lfir{L_{\rm FIR}} 
\def\ffir{F_{\rm FIR}}
\def\stwelve{S_{12\mu}}
\def\stwentyfive{S_{25\mu}} 
\def\ssixty{S_{60\mu}}
\def\shundred{S_{100\mu}}
\def\sone{S_{1.4\rm GHz}}
\def\stwo{S_{2.4\rm GHz}}
\def\sthree{S_{2.7\rm GHz}}
\def\sfive{S_{5\rm GHz}}
\def\sfirradio{s_{60\mu/1.4\rm GHz}}
\def\oneghz{1.4\, \rm GHz}
\def\twoghz{2.4\, \rm GHz}
\def\threeghz{2.7\, \rm GHz}
\def\fiveghz{5\, \rm GHz}
\def\twelvemu{12 \mu\rm m}
\def\twentyfivemu{25 \mu\rm m}
\def\sixtymu{60\, \mu\rm m}
\def\hundredmu{100\, \mu\rm m}
\def\diffe{\frac{dN_{e}}{dE_{e}}}
\def\alphain{\alpha_{e}^{\rm prim}}
\def\alphaem{\alpha_{e}^{\rm sec}}
\def\diffnu{\frac{dN_{\nu}}{dE_{\nu}}}
\def\alphanu{\alpha_{\nu}}
\def\fluxunits{\rm GeV^{-1} s^{-1} sr^{-1} cm^{-2}}
\def\diffunits{\rm GeV s^{-1} sr^{-1} cm^{-2}}
\def\pointunits{\rm GeV s^{-1} cm^{-2}}
\def\g25{\Gamma_{2.5}}
\def\lumi{L_{\gamma}^{51}}
\def\be{\begin{equation}}
\def\ee{\end{equation}}
\def\bea{\begin{eqnarray}}
\def\eea{\end{eqnarray}}
\title{Cosmic Rays VI}
\subtitle{Starburst Galaxies at multiwavelengths}
\author{Julia K.~Becker\inst{1,2}\fnmsep\thanks{Corresponding author. Contact: julia.becker@physics.gu.se, phone: +46-31-7723190} \and Peter L.~Biermann\inst{3,4,5,6,7}
          \and  Jens Dreyer\inst{2} \and Tanja M. Kneiske\inst{2,8}
          }
\institute{G\"oteborgs Universitet, Institutionen f\"or Fysik,
  41296 G\"oteborg, Sweden
\and
Technische Universit\"at Dortmund, Institut f\"ur Physik, D-44221
  Dortmund, Germany
\and
   MPI for Radioastronomy, Auf dem H\"ugel 69,
  D-53121 Bonn, Germany
         \and
Dept.~of Phys.~\&~Astron., Univ.~of Bonn,
  Germany
\and
Dept.~of Phys.~\&~Astron., Univ.~of Alabama,
  Tuscaloosa, AL, USA
\and
Dept.~of Phys.~\&~Astron., Univ.~of Alabama,
  Huntsville, AL, USA
\and
Inst.~Nucl.~Phys.~FZ, Karlsruhe Inst.~of
  Techn.~(KIT), Karlsruhe, Germany
\and
University of Hamburg, Institut f\"ur
  Experimentalphysik, Hamburg, Germany
             }
\date{\today}

\abstract
   {Starburst galaxies show a direct correlation between radio
     and far-infrared emission. High target densities and a high rate
     of supernova explosions imply the possibility of accelerating
     hadronic cosmic rays and producing decay products from hadronic
     interactions, like high-energy neutrinos and photons.}
   {We propose an explanation for the far-infrared/radio
correlation of
galaxies in terms of the energy balance of the interstellar medium and
   determine the flux from high-energy photons and neutrinos from
   starburst galaxies.}
   {We present a catalog of the 127 brightest starburst galaxies with
     redshifts of $z<0.03$. In order to investigate the correlation
     between radio- and far-infrared emission, we apply the leaky box approximation. Further, we derive photon- and
     neutrino spectra from proton-proton interactions in supernova
     remnants (SNRs). Here, we assume that a fraction of the SNR's
     energy is transferred to the acceleration of cosmic rays. We also
     investigate the possibility of detecting Gamma Ray Bursts from nearby starburst galaxies, using
     the catalog defined here.}
   {We show that the radio emission is only weakly dependent on the
     magnetic field. It turns out that the intensity of the radio signal is directly
     proportional to the number of supernova explosions, which scales
     with the far-infrared luminosity. In addition, we find that
     high-energy photons from proton-proton interactions in SNRs in starbursts can make up several percent of the diffuse
   gamma-ray background. The neutrino flux from the same sources has a maximum energy of $\sim
   10^{5}$~GeV. Neutrinos can, on the other hand, can be observed if a
   Gamma Ray Burst happens in a nearby starburst. About $0.03$ GRBs
   per year
   are expected to occur in the entire catalog. The true number is
   expected to be even higher, since we only include the brightest sources. The number of events
   per burst in IceCube varies between about one event and more than
   1000 events.  This provides good prospects for IceCube 
   to detect a significant event, since the background for a GRB
   search is close to zero.}
   {}
   \maketitle
\section{Introduction} 
\parindent=0cm
Radio emission from galaxies is usually dominated by synchrotron 
emission from a population of non-thermal, energetic electrons in a
magnetic field which permeates most of the interstellar medium.  This radio
emission is often spatially structured, such as in the starburst galaxy M82, showing
individual compact sources, which can be interpreted as fairly young 
supernova remnants \citep{kbs1985,kronberg_sramek1985,bartel1987}.  The origin of these energetic electrons, a part of 
the cosmic rays, is thus expected to be the young supernova remnants, see
\cite{baade_zwicky1934,shklovskii1953} and for a extensive 
review \cite{berezinskii1990}.  Thus the radio emission is a key to 
interpret the physics of cosmic rays, and conversely, any attempt to 
understand cosmic rays should also try to understand the properties of 
the radio emission.

Galaxies also have abundant far-infrared (FIR) emission, which is due to
dust. This
dust is heated by stars, often mostly young stars.  As was noted from the
mid-eighties, this thermal dust emission correlates rather well with the
non-thermal radio emission.  The correlation in its most simple form is
just a proportionality between far-infrared and non-thermal emission.
As reference wavelengths, $\sixtymu$ and $\hundredmu$ are used
for the far-infrared. Frequencies between $\oneghz$ and $\fiveghz$ are used
as typical for the radio regime.

Many attempts have been made to understand the proportionality between radio
and far-infrared.  On the
basis of rather simple modeling of galactic evolution, a strong
correlation is actually expected, since both supernova remnants and the dominant heating
by ultraviolet light from massive stars derives from the same stellar
population~\citep{biermann1976,biermann_fricke1977}. Using such models the
far-infrared luminosity of NGC2146 had been predicted
 by \cite{kronberg_biermann1981} and verified subsequently by IRAS observations~\citep{iras1990}.  The correlation as
seen in the data was first clearly stated by \cite{dejong1985}, and subsequently
discussed at some length by many authors \citep{bicay1989,wwk1987,wunderlich_klein1988,wunderlich_klein1991,condon1991}.  It still defies a clear explanation.

An extensive attempt to interpret the correlation was made by V{\"o}lk
and collaborators \citep{voelk1989,xu1994a,xu1994b,lvx1996a}. The electrons are believed to lose all energy in this model and
therefore, the correlation is calorimetric. However, this would
predict  an actual steepening of
 the radio emission between the reference
frequencies, 
an effect  which is not seen in the data. The solution of a spatial mixture of cutoffs would still allow for a locally loss-dominated scenario. However, first spatially resolved observations of M~33 indicate that the local star-forming regions typically have flat spectra, too~\citep{fatemeh2007a,fatemeh2007b}. It is therefore likely that star forming regions are generally injection dominated.

Here, we propose a simple model, based upon a particular picture of
the 
energy balance in the interstellar medium. This model also uses some simple
assumptions, as we will emphasize.  The model is local, and so
automatically allows for starbursts and gradients in disk galaxies, while
upholding the correlation. The model leads to some specific predictions which
can be  checked with further data. A catalog of starburst galaxies is
presented to perform first checks.

The outline of this paper is as follows: We first define variables,
used throughout the paper, in Section~\ref{definitions:sec}. In Section~\ref{catalog:sec},
a catalog of 127 nearby, bright starburst galaxies is presented, including
far-infrared, radio and X-ray data. The correlation between FIR
and radio emission is outlined together with the difficulty of explaining
it. In Section \ref{fir_radio:sec}, we present a model explaining the
FIR-radio correlation. The possible emission of cosmic rays and secondaries
produced in proton-proton and proton-photon interactions is discussed
in Section~\ref{cosmic_rays:sec}. In particular, we examine the
possibility of detecting secondaries from supernova remnants, as well
as cosmic rays and secondaries from Gamma Ray Bursts in starbursts.
Finally, implications are discussed in Section~\ref{implications}.

\section{Definitions\label{definitions:sec}}
In the following sections, the electromagnetic spectra at different wavelengths are
used in order to investigate the starburst nature of the catalog
sources. Table~\ref{parameters:tab} gives a summary of the different parameters
used. Concerning spectral power-law fits between the wavelengths, we use the convention
\begin{equation}
S=S_{0}\cdot \left(\frac{\nu}{\nu_0}\right)^{\alpha}\,,
\end{equation}
with $S$ as the flux per area and frequency interval, in units of \mbox{Jy$=10^{-26}$~W/m$^2/$Hz}. Here, $S_{0}$ is the flux at a reference frequency $\nu_0$.
\begin{table*}
\centering{
\begin{tabular}{|l|l|l|}\hline
parameter&symbol&units\\\hline\hline
Electron spectrum&&\\
--primary&$dN_{e}/dE_{e}\propto {\eel}^{-\alphain}$&$\fluxunits$\\
--secondary&$dN_{e}/dE_{e}\propto {\eel}^{-\alphaem}$&$\fluxunits$\\
Electron energy&$\eel$&keV\\
$e$ spectral index&&\\
-- primary&$\alphain$&\\
-- secondary&$\alphaem$&\\\hline
Proton spectrum&&\\
&$dN_p/dE_p=A_p\cdot (E_p/E_{\max})^{-\alpha_p}\cdot$&$\fluxunits$\\
&$\cdot\exp(-E_p/E_{\max})$&\\
Proton energy&$E_p$&GeV\\
p spectral index&$\alpha_p$&\\
p cutoff energy&$E_{\max}$&GeV\\\hline
Neutrino spectrum&$dN_{\nu}/d\en=A_{\nu}\cdot (\en/{\rm GeV})^{-\alphanu}$&$\fluxunits$\\
Neutrino energy&$\en$&GeV\\
$\nu$ spectral index&$\alphanu$&\\
normalization factor&$A_{\nu}$&$\fluxunits$\\\hline
Radio flux density&&\\
-- at $\oneghz,\,\twoghz,\,\threeghz$ \& $\fiveghz$&$\sone,\,\stwo,\,\sthree,\,\sfive$&mJy\\\hline
IR flux density, IRAS& &\\
-- at $\twelvemu,\,\twentyfivemu,\,\sixtymu$ \&
$\hundredmu$&$\stwelve,\,\stwentyfive,\,\ssixty$ \& $\shundred$&Jy\\
IR flux density, 2MASS&&Jy\\
-- at $1.25\,\mu$m, $1.65\,\mu$m \& $2.17\,\mu$m&$S_{1.25\mu\rm
  m},\,S_{1.65\mu\rm m},\,S_{2.17\mu\rm m}$&Jy\\
X-ray flux density, ROSAT&& \\
--btw.\ [0.1-4.5],[0.1-2.3] or [0.2-2.0]~keV&$S_{\rm ROSAT}$&nJy\\\hline
spectral index&&\\
-- btw $\oneghz$~\&~$\fiveghz$&$\alpha$&\\
-- btw $\oneghz$ \& $\sixtymu$&$\alpha_{\rm rir}$&\\
-- btw $\sixtymu$ \& $\sim 1$~keV&$\alpha_{\rm xir}$&\\\hline
\end{tabular}
\caption{Summary of parameters used in this paper\label{parameters:tab}.}
}
\end{table*}
\section{Local starbursts: a sample\label{catalog:sec}}
\begin{figure}
\centering{
\epsfig{file=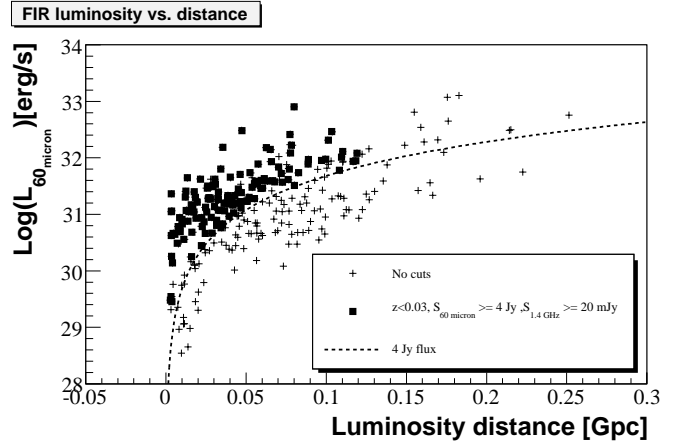,width=\linewidth}
\caption{60$\mu$ luminosity - distance diagram. Crosses indicate all
  309 pre-selected starburst galaxies, squares show those remaining
  after the cuts  $\sone>20$~mJy and $\ssixty>4$~Jy and $z<0.03$. The
  dashed line shows the sensitivity for $\ssixty>4$~Jy. \label{logl_dl}}
}
\end{figure}

\begin{figure}
\centering{
\epsfig{file=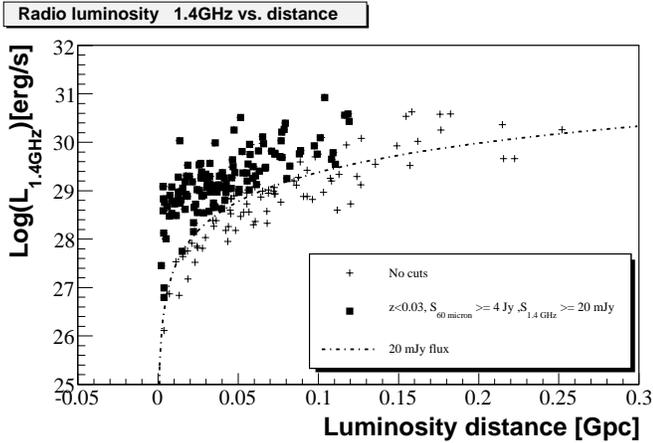,width=\linewidth}
\caption{$1.4$~GHz luminosity - distance diagram. Same notation as
  Fig.~\ref{logl_dl}.  The
  dashed line shows the sensitivity for $\sone>20$~mJy.\label{logl_dl_radio}}
}
\end{figure}

In this section, we present a sample of local starburst galaxies\footnote{The data
  have been collected from the references \citep{1964aj69_277h,1970ApL.....5...29W,1975AJ.....80..771S,1976ApJ...207..725S,1977MNRAS.179..235D,1978ApJS...36...53D,1981A&AS...45..367K,1983AJ.....88...20C,1990pks90C_0000w,1983ApJS...53..459C,1988iras....1.....B,1989AJ.....98..766S,1990irasf_c_0000m,1991apjs75_1B,1992ApJS...80..531F,1992apjs79_331w,1994A&A...281..355B,1994PrivC.U..J....K,1994ApJS...91..111W,1994ApJS...90..179G,1995ApJS...97..347G,1995apj450_559b,1996MNRAS.278.1049R,1996ApJS..103..145W,1996aj111_1945d,1996apjs103_81C,1998aj115_1693c,2000yCat.9031....0W,2002AJ....124..675C,2003AJ....126.1607S,2004ApJ...606..829S,2004A&A...418....1V,2004AJ....127.3235S,2004AJ....127..264B,2005apj625_763l,2005A&A...435..521N,2005MNRAS.358.1423O,2005ApJ...625..763L,2005ApJS..158....1I,2005A&A...435..799T,2005ApJ...633..664T,2005A&A...444..119G,2006A&A...449..559B,2006A&A...448...43T,2006AJ....132..546G,2007A&A...462..507L,2007ApJ...654..226R,2007ApJ...657..167S}
}. The data of the individual sources are presented in appendix~\ref{appendix_a}. Only local sources are
considered, since our aim is to investigate the closest sources.
The catalog presented here consists of a total of 127 starburst
galaxies. This is a sub-sample from a larger sample of 309 starburst galaxies, applying cuts at
both FIR and radio wavelengths to ensure a complete, local
sample. These cuts are discussed in the following paragraphs.
Different tests were performed in order to verify that the considered galaxies are indeed
starbursts, as presented in the following paragraphs. In order to
remove contamination from Seyfert galaxies, we only use high ratios of
FIR to radio flux density, i.e.~$\ssixty/\sone>30$. To test if the sample
consists of starburst galaxies as opposed to regular galaxies, we
check that the correlation between radio power and FIR luminosity is a
direct proportionality.
Apart from that, our main
criterion for the catalog is that the sources are closer than
$z<0.03$, i.e.~located in the supergalactic plane, and that they have
both radio and IR detections. The latter gives us information about
the ratio of the IR to radio signal, which we require to be larger
than 30. This ensures a high IR component compared to the radio part,
i.e.~that the sources are indeed starbursts and not Seyfert
galaxies. Further, we apply sensitivity cuts, we only include sources
with a flux density $>4$~Jy at $60$~$\mu$m and a radio flux density at $1.4$~GHz
larger than $20$~mJy. Figures~\ref{logl_dl} and \ref{logl_dl_radio} show the $60$~$\mu$m resp.~$1.4$~GHz luminosity of
starburst galaxies versus their luminosity
distance. The dotted lines represent the sensitivity for $4$~Jy,
resp.~$20$~mJy. Crosses represent all 309 sources we selected in the
beginning, squares show those 127 sources remaining after the cuts at
$\sone>20$~mJy and $\ssixty>4$~Jy, as well as $z<0.03$. We apply those
cuts in order to ensure a complete, local sample in both FIR and radio wavelengths.

\begin{figure}
\centering{
\epsfig{file=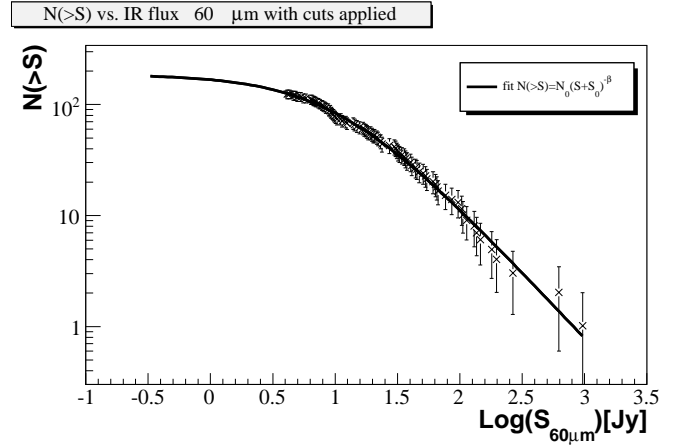,width=\linewidth}
\caption{$\log N-\log S$ representation of the catalog. An
  $S^{-1.2}-$fit matches the data nicely, with a turnover at
  $S_0=10.56$~Jy. \label{logN_logS:60mu}}
}
\end{figure}

Since the sources are closer than $z=0.03$, many of the starbursts are
located in the supergalactic plane. Their spatial distribution should
therefore be a flat cylinder with a further more spherical component,
for those sources not in the supergalactic plane. We therefore expect
that the number of sources with a flux density larger than $S$,
$N(>S)$, should follow a behavior of \mbox{$S^{-1}-S^{-1.5}$}. A pure
$S^{-1}-$behavior is expected for a flat cylinder, while a spherical
distribution results in an $S^{-1.5}-$behavior. 
Figure~\ref{logN_logS:60mu} shows the logarithmic number of sources
above an FIR flux density  $\ssixty$. We fit the data with the
following function:
\begin{equation}
N(>S)=N_0\cdot (S+S_0)^{-\beta}
\end{equation} 
Here, $N_0,\,S_0$ and $\beta$ are fit parameters. Using an error of
$\sqrt{N}$, the parameters are determined to 
\begin{eqnarray*}
N_0&=&3155\pm1297.9\\
S_0&=&(10.56\pm3.78)\,\rm{Jy}\\
\beta&=&1.2\pm0.2\,.
\end{eqnarray*}
The behavior \mbox{$N(>S)\sim S^{-1.2\pm 0.2}$} matches the expectation that
the function should lie between $S^{-1.0}$ and $S^{-1.5}$.
In the following paragraphs, we will investigate further whether the
classification of the 127 sources as starbursts is justified.
\subsection{FIR luminosity versus Radio power}
Looking at a well defined sample of galaxies, it turns out that the 
correlation between radio and far-infrared (FIR) emission is {\it not} linear, i.e., that the radio luminosity is 
proportional to the far-infrared luminosity to the power \mbox{1.30
  $\pm$ 0.03} \citep{xu1994a}. As Xu and collaborators note, the far-infrared
emission 
has two heating sources, stars that do explode later as supernova
remnants, 
and also stars, that will never explode as supernovae. This second
population 
of stars needs to be corrected for, and their contribution to the dust 
heating needs to be eliminated.  This then leads to a corrected far-infrared
luminosity,  which is directly proportional to the radio luminosity \citep{xu1994a}.

The proportionality holds along a disk in a galaxy, even for
fairly 
short lived phases like a starburst, such as in M~82, and thus requires clearly
local physics, with a short readjustment time scale.  This poses a severe
difficulty for any proposal to explain the radio/FIR correlation.

\begin{figure}
\centering{
\epsfig{file=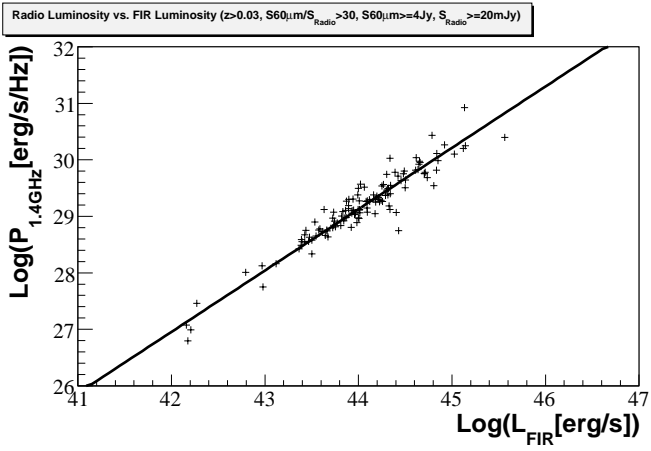,width=\linewidth}
\caption{Radio power $P_{1.4\rm GHz}$ at $\nu=1.4$~GHz versus FIR
  luminosity $\lfir$. A direct proportionality,
  $P_{1.4\rm GHz}\propto L_{FIR}$ is found.\label{fir_radio}}
}
\end{figure}

The FIR luminosity in the range of $\sixtymu$ and $\hundredmu$ is given as~\citep{xu1994a}:
\begin{equation}
\lfir:=4\,\pi\,d_{l}^{2}\cdot \ffir\,.
\end{equation}
Here, $d_{l}$ is the luminosity distance of the individual sources and 
\begin{equation}
\ffir:=1.26 \cdot 10^{-14}\cdot\left[2.58\cdot \left(\frac{\ssixty}{\rm Jy}\right)+\left(\frac{\shundred}{\rm Jy}\right)\right]{ \rm W\,m}^{-2}
\end{equation}
is the FIR flux density at Earth as defined in~\cite{helou1988}. The normalization factor comes from the frequency integration and from the conversion of Jy to W/m$^2$/Hz. 
In Fig.~\ref{fir_radio}, the logarithm of the radio
power at 1.4~GHz, $P_{1.4\rm GHz}$ versus
the logarithm of the FIR luminosity $\lfir$ is shown for our catalog.
The circles show the single sources and the solid
line is a fit through the data. The fit yields a correlation of
\begin{equation}
P_{1.4\rm GHz}\propto {\lfir}^{1.0}\,.
\end{equation}
This demonstrates that short stellar lifetimes dominate the correlation
in our sample, and so this is strongly supporting our hypothesis, that
the majority of our sample galaxies are starbursts.

\subsection{Infrared to radio flux density ratio}
\begin{figure}[h!]
\centering{
\epsfig{file=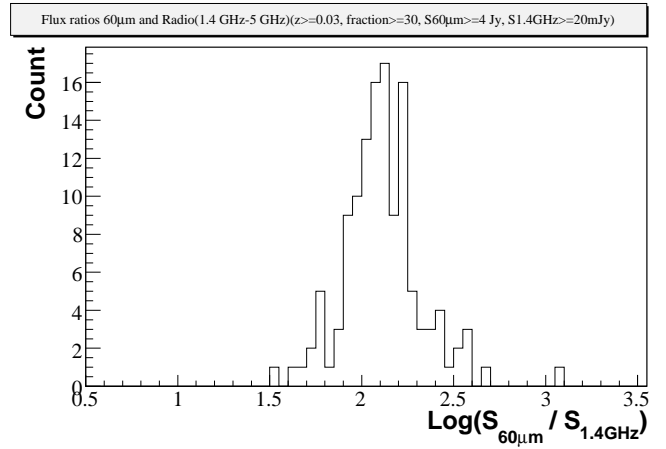,width=\linewidth}
\caption{Ratio of the flux density at $60\,\mu$m and at $1.4$~GHz. All sources in the
  sample have ratios larger than $30$, which indicates a high star formation
  rate. The median is around $100$. This matches previous
  investigations, e.g.
  \cite{biermann1985}, who find a mean value of 250 at higher radio frequencies, $\nu=5$~GHz. \label{60_mu_radio}}
}
\end{figure}

Generally, regular galaxies are distinguished from active galaxies by their ratio
of the FIR flux density at $\sixtymu$, $\ssixty$, and the radio flux
density at $\oneghz$, $\sone$:
\begin{equation}
\sfirradio:=\frac{\ssixty}{\sone}\,.
\end{equation}
For Seyfert galaxies, this ratio is about
\mbox{$\sfirradio\sim 10$}, while it is significantly higher in the case
of starburst
galaxies, \mbox{$\sfirradio \sim 300$}. 
 The histogram of the ratio between the FIR flux density at $\sixtymu$
 and the radio flux density at $\oneghz$ is shown in
Fig.~\ref{60_mu_radio}. All 127 sources 
have a ratio of \mbox{$\sfirradio>30$}, which confirms that the sources
are not likely to be Seyferts.
\subsection{Radio to Infrared and X-ray to Infrared spectral indices}
\begin{figure}[h]
\centering{
\epsfig{file=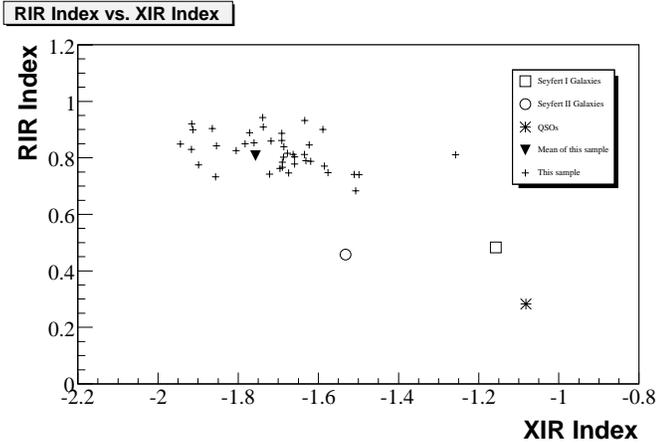,width=\linewidth}
\caption{Radio-to-IR spectral index versus X-ray-to-IR spectral
  index. The crosses represent those 48 sources in our catalog with
  radio, FIR and X-ray measurements. The blue triangle shows the
  average of the values. The open circle shows the average location of Seyfert-I
  galaxies, the open square represents average Seyfert-II galaxies and the
  star indicates QSOs. The last three values are taken from
  \cite{rodriguez_pascual1993}. Note that individual galaxies scatter
  around the given values \citep{chini1989}.\label{rir_xir}}
}
\end{figure}
A further criterion of distinguishing regular galaxies and Seyferts is
their spectral index from X-ray to IR (XIR) and from radio to IR (RIR). The diagram of
the XIR (1~keV to 60~$\mu$m) versus RIR (5~GHz to 60~$\mu$m) index of the sources in shown in
Fig.~\ref{rir_xir}. Derived from figure 3 in
\citep{rodriguez_pascual1993}, starburst galaxies have spectral
indices scattering around \mbox{(RIR,
XIR)$_{starburst}\sim(0.6,\,-1.9)$}, Seyfert-I galaxies show \mbox{(RIR,
XIR)$_{Sy-I}\sim(0.48,\,-1.2)$}, Seyfert-II galaxies have \mbox{(RIR,
XIR)$_{Sy-II}\sim(0.47,\,-1.6)$} and quasars are located at \mbox{(RIR,
XIR)$_{quasar}\sim(0.28,\,-1.1)$}. The values for the RIR and XIR indices of starburst galaxies given
by \cite{chini1989} are slightly higher, which matches the sample
examined here: \cite{chini1989} give a RIR index of $0.82$ and a XIR
index of $-1.66$. We find average values of \mbox{$(\overline{{\rm RIR}},\,\overline{{\rm
    XIR}})=(0.82,\,-1.77)$} which is compatible with the expected
result. 

Still, we do not have X-ray data for all the sources, so there may
still be some contamination from both Seyferts and regular galaxies in
the sample. As we only used catalogs where the sources have previously
been identified as starbursts, this contamination should be small.

\section{The interstellar medium and the FIR/radio correlation \label{fir_radio:sec}}
The interstellar medium connects the formation of stars, the explosion 
of supernova remnants, the regularization and enhancement of the magnetic 
field, and the transport of cosmic rays. In order to understand the 
observation that the thermal hot dust emission from a galaxy is simply 
proportional to the non-thermal radio emission from relativistic cosmic 
ray electrons, we need to understand the interstellar medium, or at 
least get close enough to a comprehension, that we can understand this 
amazing correlation
\citep{dejong1985,wwk1987,wunderlich_klein1988,wunderlich_klein1991,condon1991}.

Obviously, since the very massive stars power the far-infrared emission 
through a large fraction of the ultra-violet emission, which is absorbed 
by dust, and then in supernova explosions produce the energetic cosmic 
ray electrons, which emit the observed radio emission, there should be a 
correlation. Using simple initial models for stellar population 
evolution, this rather naive early picture successfully predicted the 
approximate far-infrared emission already in 1977
\citep{biermann1976,biermann_fricke1977,kronberg_biermann1981}. However, the tightness 
of the correlation was neither anticipated nor predicted. The 
correlation was finally discovered by \cite{dejong1985}. Since the 
dust emission just measures the total output in ultra-violet by young 
stars, it seems obvious, that in the limit of much absorption, the 
far-infrared emission would just be proportional to the star formation 
rate, given a general initial mass function. However, the radio emission 
is approximately proportional to the product of the cosmic ray electron 
density and the magnetic field energy density, and therefore it is not 
really obvious at all, that integrating along a vertical column through 
the disk of a galaxy these two emission components should be basically 
proportional.

Modern descriptions of starburst galaxies such as \mbox{M 82} or \mbox{NGC 2146} are in \cite{dopita2005,dopita2006a,dopita2006b,groves2008}.

It had been observed early that the three main components of the 
interstellar medium, the gas, the cosmic rays, and the magnetic field 
have very similar energy densities, or pressures. Since all three derive 
from very different physical processes, to keep them at approximate 
equipartition implies that the three time scales of change are also all 
three the same. This is the basic premise of the following argument, and 
it will lead naturally to an understanding of the far-infrared radio 
correlation. So in this specific sense it is a calorimetric argument 
similar to \cite{voelk1989}, although we approach the problem is a 
somewhat different way.

The following line of reasoning is visibly influenced by earlier papers, 
like \cite{biermann1950,biermann_schlueter1951,cox1972,cox_smith1974,mckee_ostriker1977,beuermann1985,kbs1985,kronberg_sramek1985,snowden1997,hunter1997,beck2003,hanasz2004,hanasz2006}, and of course others 
mentioned in due course.

First we wish to establish the concepts which we use, for easy 
reference, and then apply them to the problem here.
\subsection{The three main components of the interstellar medium}
The three main components are the gas, the cosmic rays and the magnetic 
field. At least the gas and the magnetic field is clearly spatially highly 
inhomogeneous\citep{beck2003}:

The gas has a number of components, molecular clouds, neutral Hydrogen 
clouds, diffuse neutral Hydrogen, diffuse ionized Hydrogen, HII regions, 
stellar wind bubbles, supernova remnants with X-ray emitting shells, a 
tunnel network of connected older supernova remnants \citep{cox_smith1974}, a thick hot disk \citep{beuermann1985,snowden1997,kronberg2007}, and a wind \citep{breitschwerdt2008,everett2008}. The tunnel network probably connects to the hot thick disk. The 
wind is probably fed from the hottest regions of the tunnel network, in 
a fashion perhaps similar to the Solar wind being fed from coronal holes \citep[e.g.]{stepanian2008}.

The magnetic field is permeating almost everything, and also has a thick 
disk. The field permeates the clouds, and is often confined visibly by 
the clouds \citep[e.g.]{appenzeller1974}. The magnetic field is strongly 
perturbed by HII regions, and supernova explosions. The magnetic field 
is transported out of the disk by the wind.

The cosmic rays, produced by supernova explosions in shockwaves
\citep{baade_zwicky1934,fermi49,drury1983,berezinskii1990} cannot 
easily be repelled by anything, and so go through all clouds, and all 
the neutral and ionized gas. In bulk they cannot travel faster then the 
Alfv{\'e}n speed, since otherwise they would excite waves in the plasma, 
scattering the particles, effectively reducing their bulk velocity.

The magnetic field is instrumental to allow cosmic ray acceleration in 
shocks, and perhaps throughout the medium.

The cosmic rays in turn drive the dynamo mechanism to enhance an existing 
magnetic field and give it spatial coherence \citep{parker1969,parker1992,ferriere1996,ferriere_schmitt2000,hanasz2004,hanasz2006,otmianowska_mazur2008}. In the classical Biermann-battery 
mechanism \citep{biermann1950,biermann_schlueter1951}, only a 
rotating star with surfaces of density and pressure in non-coincidence 
is required to produce a seed field, which, however, is generally weak; 
the dynamo mechanism in stars can strongly enhance magnetic fields, and 
through winds eject them \citep[e.g.]{bisnovatyi_kogan1973}: This 
would constitute a very irregular, but potentially relatively strong 
source of magnetic fields in galaxies; in such a case the dynamo 
mechanism on a Galactic scale is required more to regularize the field 
rather than to strengthen it. In the mechanism of \cite{lucek_bell2000,bell2004,bell2005} the shock waves can directly enhance the magnetic 
field, using an existing population of cosmic rays. As noted already, 
the cosmic rays couple effectively to the gas. Anisotropies in the phase 
space distribution of particles, always present in shockwaves, also can 
produce new magnetic fields on small scales \citep{weibel1959,bykov_toptygin2005}. Galactic magnetic fields have also been reviewed by
\cite{beck1996,kulsrud1999,kulsrud_zweibel2008}.

Thermodynamically both the cosmic rays and the magnetic field can be 
thought of as a relativistic gas, with almost zero net mass density.

Therefore the ensemble of cosmic rays and magnetic field constitute a 
light fluid pushing against the heavy fluid of the normal gas, and given 
enough energy density, these two components escape via an instability \citep{parker1965,kowal2003,kowal2006}.

Therefore all three components are strongly coupled, and the data 
confirm an approximate energy equipartition between cosmic rays and the 
magnetic field, and the sum of these two components equal in energy 
density to the gas.

The one given parameter is the total energy input, integrated across all 
spatial inhomogeneities, since the energy supply is given by the stars, 
in the form of winds, explosions, and radiation. Since the 
inhomogeneities are extreme, especially in the density, it is important 
to use spatially integrated energy densities for reference as much as 
possible. So we note that the energy density of the magnetic 
irregularities integrated over all spatial scales is also in approximate 
equipartition \citep{beck1996}.

The energy input can be estimated from the explosions of supernovae to 
about 1 supernova of $10^{51}$~erg, every 100 years, so at \mbox{$L_{kin} \; 
= \; 3 \cdot 10^{41} \, {\rm erg/s}$}; the uncertainty in this is about a 
factor of 3. Some fraction of this energy goes into cosmic rays. This 
fraction could be large \citep{drury1983}.

Other energy input can be estimated from the infrared emission \citep{cox_mezger1989}: About 1/3 to 1/4 of the total stellar radiation is 
absorbed by dust and reradiated in the infrared, beyond a wavelength of 
25 $\mu$, about \mbox{$10^{10} \; L_{\odot}$}. Of this, about \mbox{$2 \cdot 10^{9} 
\, L_{\odot}=8 \cdot 10^{42} \; {\rm erg/s}$} is coming from young star forming regions.

The energy density of magnetic fields can be estimated to be about \mbox{$1.6 
\cdot 10^{-12}$ dyn/cm$^2$} \citep{beck1996,everett2008}, the energy density of cosmic rays is about the 
same, and their sum is about equal to the gas pressure, of \mbox{$4 \cdot 
10^{-12}$ dyn/cm$^2$}. Using a scale height of full width of 3 kpc, and a 
radius of 10 kpc, we obtain a crude estimate of the energy content. This 
requires for magnetic fields and cosmic rays together an average supply 
of energy of about \mbox{$4 \cdot 10^{41} \, {\rm erg/s}$}, using the time 
scale obtained from cosmic rays (see below). Such numbers are uncertain 
by probably a factor of 2. \cite{everett2008} also estimate the 
required wind-power to about \mbox{$4 \cdot 10^{41} \, {\rm erg/s}$}, a Galactic 
wind driven by cosmic rays; they discuss other estimates.

It is interesting to note that to within the uncertainties all these 
power estimates (supernovae, wind power, magnetic field and cosmic ray 
replenishment) agree better than their respective error estimates.

Therefore the time scales to replenish anyone of the components must 
also be approximately be the same. We do have the real number from 
radioactive isotopes of cosmic rays interacting, and the number is about 
10 million years \citep[e.g.]{brunetti_codino2000}. This is then the time 
scale for all key processes.
\subsection{Supernova explosions}
For didactic simplicity we basically adopt the approach of \cite{sedov1958,cox1972}, but use more modern cooling approximations, and allow 
for much lower environmental densities, but otherwise rescale their equations.

An explosion runs into the interstellar medium, and expands into the 
tenuous gas, which surrounds the clouds, and extends far above and below 
the central layer of cool clouds. We consider the expansion into the 
surrounding low density medium and ask, when the expansion runs into the 
cooling limit:

The first question is what density should be used: The galaxy has a
wind \citep{westmeier2005,breitschwerdt2008,everett2008,gressel2008,otmianowska_mazur2008} and it is getting fed from 
the tunnel network of \cite{cox_smith1974} probably. This implies that 
the Alfv{\'e}n speed must approach the escape speed, for a cosmic ray 
driven magnetic wind. Other galaxies also show evidence for winds \citep{chyzy2000,chyzy2000_err,chyzy2004,chyzy2006,chyzy2007}:
\begin{equation}
V_A \; = \; \frac{B}{4 \pi \rho} \; \simeq \, 400 \, {\rm km/s}\,.
\end{equation}
With \mbox{$B \approx \, 3 \, \mu{\rm Gauss}$}, this implies a density of about 
\mbox{$n \, = \; 3 \cdot 10^{-4} \, {\rm cm^{-3}}$}.

The time to start cooling is
\begin{equation}
\tau^{c} \; = \; 5 \cdot 10^{6} \, {\rm yrs} \, 
{\left({\frac{E_{51}}{n_{-3.5}}} \right)}^{2/11} \, {\left( {r 
\Lambda_{-21} \, n_{-3.5}}\right)}^{-5/11}\,,
\end{equation}
where $E_{51}$ is the energy of the explosion in units of \mbox{$10^{51}$ 
erg}, $n_{-3.5}$ is the tenuous density in units of \mbox{$ 3 \cdot 10^{-4}$ 
cm$^{-3}$}, $\Lambda_{-21}$ is the cooling coefficient in units of 
\mbox{$10^{-21}$ erg cm$^{3}$ s$^{-1}$}, and $r$ is a compaction parameter of 
order unity. This low density reflects the finding that the tenuous 
medium is of very low density, and due to substructure may on volume average be 
of even lower density than suggested by the X-ray data
\citep{snowden1997,everett2008}, of order \mbox{$3 \cdot 10^{-3} \; {\rm cm^{-3}}$}; 
but we do use the temperature of \mbox{$10^{5} \; {\rm K}$}, near the maximum, 
and also close to the stable region \citep{field1965}. However, the cooling 
suggested by the X-ray spectrum is an integral over the entire 
evolution, and so only sensitive to the earliest part of the evolution. 
As soon as two or more supernova remnants overlap, and start building a 
network \citep{cox_smith1974}, then the temperature evolution will be 
different, giving again higher temperatures, consistent with observations.

The radius at that stage is
\begin{equation}
R^{c} \; = \; 8 \cdot 10^{2} \, {\rm pc} \, 
{\left({\frac{E_{51}}{n_{-3.5}}} \right)}^{3/11} \, {\left( {r 
\Lambda_{-21} \, n_{-3.5}}\right)}^{-2/11}
\end{equation}
and the temperature then is initially
\begin{equation}
T^{d} \; = \; 1.7 \cdot 10^{5} \, {\rm K} \, 
{\left({\frac{E_{51}}{n_{-3.5}}} \right)}^{2/11} \, {\left( {r 
\Lambda_{-21} \, n_{-3.5}}\right)}^{6/11}\,.
\end{equation}
This corresponds to an injection scale of turbulence. 
Interestingly, the time scale is of the same order of magnitude to what 
we derive from cosmic ray transport, and the length scale is not far 
from the scale height of the hot disk \citep{snowden1997}, 
demonstrating qualitative consistency. We note that \cite{snowden1997} gave 
a much higher temperature, of about \mbox{$4 \cdot 10^{6}$ K}, with a density of 
\mbox{$3 \cdot 10^{-3} \; {\rm cm^{-3}}$}. Also, \cite{everett2008} suggest a 
higher temperature. However, the luminosity of that phase is a very 
small fraction of the entire dissipation in the ISM.

As \cite{field1965} shows the cooling is stable if the temperature 
dependence of the cooling function $\Lambda$ is sufficiently strong and 
its double logarithmic derivative positive, or in the presence of 
heating larger than 2. This is the cooling phase we are interested in. 
This is the case at temperatures below and near about \mbox{$\sim 10^{5}$ K}.

Data suggest that the break-up of supernova remnant shells has been 
observed in the starburst galaxy M82 \citep{bartel1987}. However, in 
that case it is not clear whether we are observing the break-up of a 
wind-shell produced in a snow-plow effect by the stellar wind prior to 
the supernova explosion, or the break-up of the snow-plow of the normal 
supernova exploding into the interstellar medium.
\subsection{Magnetic inhomogeneities}
In the magnetic field data in our galaxy \citep{beck2003} there is 
already strong evidence for small scale substructure, since different 
measures of the magnetic field yield very different numbers: 
{lin\-e\-ar} measures such as Faraday Rotation Measures indicate much 
lower strengths of the magnetic field than quadratic measures such as 
synchrotron emission. This is typical for small scale substructure \citep{lee2003,deavillez_breitschwerdt2004,deavillez_breitschwerdt2007}, where for a given 
total energy content high intensity sheets can hold all the energy for a 
small volume fraction. In such a picture linear measures give a much 
smaller number than quadratic measures, as is well known from 
mathematically isomorphic arguments in thermal emission. Of course we 
should be comparing the proper integrals, also involving the spatial 
distribution of thermal electron density and cosmic ray electron density 
(see, e.g., \cite{bowyer1995}). We ignore all this in our simple exercise.

We can quantify this by integrating along a long thin cylinder of unit 
length. We refer to the magnetic field as $B_{0}$, when it is 
homogeneous, and for the inhomogeneous case the magnetic field is $B_{1}$ 
over most of the length, and enhanced by a factor $1/x$ in a region of 
length $x$: This then gives for the integrated energy density

\begin{equation}
B_{1}^{2} \cdot \frac{1}{x} + B_{1}^{2} \cdot (1 - x) \; = \; B_{0}^{2}\,.
\end{equation}

\noindent Keeping the integrated energy content $B_{0}^{2}$ constant, 
the linear measure of the magnetic field is given by

\begin{equation}
B_{1} \cdot \frac{1}{x} \cdot x + B_{1} \cdot (1 - x) \; = B_{1}
\cdot (2 - x)\,.
\end{equation}

Combining the two expressions gives

\begin{equation}
\sqrt{\frac{x}{1 - x}} \cdot (2 - x)
\end{equation}

for the ratio of linear measure versus quadratic measure. In the limit 
of small $x$ this is just $\sqrt{x}$. The observations suggest that this 
ratio is of order 1/5 \citep{beck2003}, and so $x = 0.04$ by order of 
magnitude. This implies that most of the magnetic energy is contained in 
shells of a volume a few percent, possibly as low as 1 percent. Since 
the linear measure is proportional to the bending of ultra high energy 
cosmic rays, this implies that the bending is reduced by a factor 
between 5 and 10 over what we might reasonably expect otherwise. 
Obviously, in realistic situations much of this effect will be smoothed 
out, and so perhaps even more extreme situations may be required.

Using the approach of \cite{cox1972} with the environment of the tenuous hot 
phase of the interstellar medium \citep{snowden1997,everett2008} the cooling stage of an expanding shell of a supernova remnant 
might lead to such a configuration, of a very thin shell at large 
distances, with strong magnetic fields. In such a picture this stage 
would encompass most of the supernova's energy dissipation, and so 
similar considerations may apply to the interpretation of the X-ray data 
\citep{snowden1997,everett2008}.
\subsection{Turbulence}
Turbulence is an ubiquitous phenomenon, and also is a key ingredient in the
interstellar medium (see reviews by \cite{rickett1977,grm1995}).
Key concepts to turbulence theory have been introduced by \cite{prandtl1925,
vonkarman_howarth1938,kolmogorov1941a,kolmogorov1941b,kolmogorov1941c,obukhov1941,heisenberg1948,kraichnan1965}, and have been
reviewed by \cite{sagdeev1979}. One key argument which we wish to use, is the concept
of the
turbulent cascade. There the energy of the turbulence is injected into
the
gas at some large wavelength, and cascades down through wavenumber
space,
to the small wavelengths where the energy is dissipated. In many
examples
this leads in a three-dimensional isotropic model to the Kolmogorov cascade,
which can be described in a local approximation by the following diffusion
equation in wavenumber space \citep{mcivor1977,achterberg1979}:

\begin{equation}
{{d} \over {dt}} \, {{I(k)} \over {4 \pi k^2}} \, - \, {1 \over k^2} \,
{\partial \over {\partial k}} \left( {k^4 \over {3 \tau_k}} \,
{\partial \over {\partial k}} \left( {{I(k)} \over {4 \pi k^2}}
\right) \right) \; = \; A \, \delta (k - k_o)
\end{equation}

Here $I(k)$ is the energy density of the turbulence per wavenumber $k$,
and
per volume element, and $\tau_k$ is the time scale of diffusion, which
can
be written as

\begin{equation}
\tau_k \; = \; {1 \over {k {\left( \gamma_{eff} I(k) k / \rho 
\right)}^{1/2}}}.
\end{equation}

Here $\rho$ is the matter density, and $\gamma_{eff}$ is an effective
adiabatic constant for the turbulent energy. The turbulence has a
source-term, here limited to a single wavenumber $k_o$. The turbulence
diffusion equation basically says that the turbulence moves through
wavenumber space with no additional source or sink, as a constant energy
current in wavenumber phase space \citep{kolmogorov1941a,kolmogorov1941b,kolmogorov1941c}. The
solutions to this diffusion equation can be written as

\begin{eqnarray}
&& I(k) \; \sim \; k^2 \; \; {\rm for} \; \; k \le k_o \; {\rm and} \\
&& I(k) \; \sim \; k^{-5/3} \; \; {\rm for} \; \; k \ge k_o \nonumber.
\end{eqnarray}

This latter behavior is commonly referred to as the Kolmogorov cascade,
and is
found ubiquitously in nature.

\subsection{The cooling of the interstellar medium}
The interstellar medium has a number of phases, which appear to be in approximate pressure equilibrium. Concentrating on the phase of the highest
temperature, the highest speed of signalling (be it sound waves, or 
Alfv{\'e}n
waves, or other wave modes), we note that its temperature is in the
range where
the typical gaseous emission is detected in the X-ray regime. The cooling
curve of such a gas has been extensively discussed by many \citep[e.g.]{cox1972,sutherland_dopita1993,dopita_sutherland2003}.
It shows the following features, starting at low temperature, and 
considering
the cooling coefficient $\Lambda (T)$ with $n$ the interstellar density
in \mbox{$\rm
particles \; cm^{-3}$}
\begin{equation}
n^{2} \, \Lambda (T) \; \rm erg/cm^3/sec \,.
\end{equation}
This cooling curve $\Lambda (T)$ rises from near \mbox{$10^4$ K} to a local
peak near
\mbox{$\approx 3 \cdot 10^{5}$ K}. The level of cooling along this local peak is
given by
\mbox{$\Lambda \, \approx \, 10^{-21} \, \rm \rm ergs/cm^{-3}/s$}, dropping 
to \mbox{$\Lambda \, \approx \, 10^{-22} \, \rm \rm ergs/cm^{-3}/s$} near 
\mbox{$\approx 10^{6}$ K}, and towards a minimum near \mbox{$\Lambda \, \approx \, 3 
\cdot 10^{-23} \, \rm \rm erg/cm^{-3}/s$} in the range \mbox{$\approx 3 \cdot
10^{6}$ K} to \mbox{$\approx \, 10^{8}$ K}. The
peak is
due to many edges and lines in the soft X-ray range. At higher temperatures
the continuum emission begins to dominate and that emission is then
given by
\mbox{$\Lambda \, \simeq \, 1.4 \cdot 10^{-27} \, T^{1/2}$ $\rm erg/cm^{-3}/s$}. 
The sharp
cutoff to low temperatures near \mbox{$10^4$ K} is due to beginning
recombination and
thus a smaller density of free electrons to interact with. A contemporary
discussion including the effects of strong departures from ionization
equilibrium has been given by \cite{schmutzler_tscharnuter1993,breitschwerdt_schmutzler1994}.

\subsection{The leaky box approximation}
Consider a column perpendicular to a galactic disk of height $H$, and
the
number of cosmic ray particles in it as a function of particle energy
and
time $N(E,t)$; then we have the balance equation

\begin{equation}
{d \over {dt}} N(E,t) \, + {N(E,t) \over {\tau (E)}} \; = \; Q(E).
\end{equation}

\noindent where $\tau (E)$ is the escape time scale. The sign of the second
term is positive, since the process described is a loss. In a
stationary
state we then obtain readily
\begin{equation}
N(E) \; = \; Q (E) \, \tau (E).
\end{equation}
The characteristic time of loss can be written as
\begin{equation}
\tau (E) \; \sim \; {H^2 \over \kappa}\,.
\end{equation}
Here, $\kappa$ is the diffusion coefficient of cosmic ray particles,
which can
be written approximately in the quasi-linear approximation as
\begin{equation}
\kappa \; \sim \; {1 \over 3} \, r_g \, c \, {{B^2 / 8 \pi} \over {I(k) k}}\,,
\end{equation}
where we write for the turbulent energy density
\begin{equation}
I (k) k \; \sim \; I_o k_o (k/k_o)^{-2/3} \; \sim \;
I_o k_o B^{-2/3} r_o^{-2/3} E^{2/3}\,,
\end{equation}
using here the assumption of a Kolmogorov spectrum. Here,
$r_g \sim 1/k \sim E/B$ is the Larmor radius of a particle of energy $E$ 
under
consideration, which gyrates in a magnetic field of strength $B$. The basic
radius $r_o \sim 1/k_o$ corresponds to the injection scale of the
turbulence.
This then leads to a dependence of the diffusion coefficient on the various
parameters
\begin{equation}
\kappa \; \sim \; E^{1/3} B^{5/3} r_o^{2/3} I_o^{-1} k_o^{-1}\,.
\end{equation}
Therefore, we adopt the point of view that the leakage time scale
$\tau(E)$ is proportional to (relativistic) energy $E^{-1/3}$, and so
that
the equilibrium density of energetic particles $N(E)$ is proportional to
$E^{-1/3}$ as well.

There is a difficulty with this argument, which can be solved: The 
secondary to primary
ratio of
cosmic ray nuclei such as the ratio Boron to Carbon already give
information
as to the energy dependence of the leakage time, and such an analysis
gives
an energy dependence as $E^{-0.6 \pm 0.1}$ \citep[e.g]{engelmann1990}.
We
have argued elsewhere already \citep{biermann1995,wiebel1995,wiebel1998}, that
this reflects the energy dependence of the amount of target material
seen for
spallation. Most of the target interaction with heavy nuclei among the 
cosmic rays happens near the most massive stars, the Wolf-Rayet stars, 
with the turbulence excited by the cosmic rays themselves, giving an 
energy dependence of the spallation secondaries of $E^{-5/9}$
\citep{biermann1998,biermann2001,biermann2006}, consistent with the data, 
which give $E^{-0.54}$ \citep{ptuskin1999}. On the other hand, most of 
the gamma emission from $\pi$-zero decay arises from the interaction 
among the more numerous supergiant stars, the red supergiants, for which 
we suggest that the turbulence arises from instabilities.
\subsection{The radio emission}
In order to calculate the radio emission from the total number of 
relativistic
electrons in a column we first have to proceed to work out the cosmic
ray
loss time for electrons, second the cooling time for the tenuous hot
medium,
put them equal, and third calculate the radio emission per supernova
event
from a column in the disk.

The maximum energy in the elemental distribution and spectrum of the cosmic
rays is in protons near their rest mass energy, since their spectrum is 
steeper
than 2 in energy. This means that the energy which has to be used in the
expression for the leakage time is fixed. So, the leakage time is given by
\begin{equation}
\tau_{CR} \; \sim \; {{H^2 I_o k_o} \over {B^{5/3} r_o^{2/3}}\,E^{1/3}}.
\end{equation}

The maximum energy content of the population in the protron spectrum is 
here near $E = 1$~GeV, so that
\begin{equation}
\tau_{CR}(E=\rm{1\,GeV}) \; \sim \; {{H^2 I_o k_o} \over {B^{5/3} r_o^{2/3}}}.
\end{equation}
in this case.

The cooling time is given by
\begin{eqnarray}
\tau_{cool} \; && \sim \; {{n k_B T} \over {n^2 \Lambda (T)}} \; \sim
\; {{B^2/8 \pi} \over {\left(B^2/8 \pi /2 k_B T_{\star} \right)}^2} \,
{1 \over
{ \Lambda (T_{\star})}} \; \label{taucool}\\
&& \sim \; \frac{T_{\star}^{2}}{\Lambda (T_{\star})} \, {1 \over B^2} \; 
\sim \; {1 \over B^2} \nonumber
\end{eqnarray}
\noindent in the approximation that \mbox{$\Lambda (T_{\star})/T_{\star}^2 $} is
a constant, and assuming equipartition again. This is reasonable in the 
dissipation stage of supernova remnants when $\Lambda$ approaches a 
double-logarithmic derivative of 2, below a temperature of $10^{5}$ K. 
Also, in that temperature range the cooling time is a minimum, for a 
given energy density. We noted above that the dissipation stage reaches 
those temperatures.

Putting the two time scales equal then yields the relation
\begin{equation}
{{H^2 I_o k_o} \over {B^{5/3} r_o^{2/3}}} \; \sim \; {1 \over B^2} .
\end{equation}

Now, we wish to consider electrons, which emit radio emission at a certain
frequency $\nu$, which gives the condition that
\begin{equation}
E \; \sim \; \nu^{1/2} B^{-1/2}
\label{nu_c}
\end{equation}

and so, making use of the frequency and magnetic field dependence of
Equ.~(\ref{taucool}) and Equ.~(\ref{nu_c}), the diffusion time for 
electrons which emit at $\nu$ is
given by
\begin{equation}
\tau_{CR,e} \; \sim \; H^2 E^{-1/3} B^{-5/3} r_o^{-2/3} I_o k_o \; \sim \;
B^{-11/6} \, \nu^{-1/6}
\end{equation}

The emission is from a spectrum of electrons of $E^{-2.42 \pm 0.04}$ at
injection \citep{biermann_strom1993}, and so the total radio emission 
$S_{\nu}$
per supernova event in a column in a disk can be written as
\begin{eqnarray}
S_{\nu} && \sim B^{1.71 \pm 0.02} \, E_{SN,CR,e} \, \nu^{-0.71 \pm
0.02} \,
B^{-11/6} \, \nu^{-1/6} \\ && \sim \; B^{-0.12 \pm 0.02} \, \nu^{-0.88
\pm
0.02} \nonumber \end{eqnarray}
\noindent for the case that all supernova remnants produce the same
number of
relativistic electrons. Here, we use an integration along a vertical 
column to obtain the total number of energetic electrons per supernova 
explosion. This is reasonable since we are using the adiabatic
phase of supernova remnant evolution, where the energy is conserved. 
This argument leads to such a weak dependence on the strength of the 
magnetic field, that the resulting offset is smaller than the errors in 
the data.

Therefore the radio emission is only very weakly dependent on the magnetic
field, independent of all other parameters, and is directly proportional
to the
number of supernovae per time interval, and so also to the luminosity of
massive stars.

The cosmic ray electron data show that above a few 10 of GeV the 
spectrum is already in the loss limit, so steeper by unity \citep{kardashev1962}. Below that energy the direct data are compromised by the Solar 
wind modulation, but from the radio emission of normal galaxies we do 
know that the electron spectrum is closely in agreement with what we 
obtain for protons, at somewhat higher energy, allowing us to conclude 
that the electron spectrum corresponds to the diffusion limit, when the 
leakage out of the Galaxy is faster than the synchrotron loss. Using the 
expressions for the two time scales we can check:

Radio emission at 5 GHz corresponds to an electron energy of 8 GeV, and 
a synchrotron loss time of \mbox{$4 \cdot 10^{7}$ yr}.

As we note elsewhere, galaxies and starburst galaxies are sometimes so
young as to be in the injection limit, or so old as to be in the loss
limit, rather than the leakage or diffusion limit that we consider
here.  For any reasonably young age of a starburst there is always some
low radio frequency $\nu_1$, below which we are still in the injection
stage, below which the age of the starburst is longer than the diffusion
time scale.  So below that the radio flux density is lower than the
diffusion limit derived above, and the radio flux density has the
spectrum of injection, following \cite{biermann_strom1993} \mbox{$S_{\nu} \,
\sim \, {\nu}^{-0.71}$}, and so relative to the diffusion limit the radio
emission at frequency \mbox{$\nu \, < \, \nu_1$} is weaker by the factor
$(\nu/\nu_1)^{1/6}$; this is typically not far below unity.  At the the
other extreme, considered by \cite{voelk1989} there is also a radio
frequency $\nu_2$, beyond which the synchrotron and inverse Compton
losses cut in, and become faster then the diffusion.  This is directly
visible in the observed cosmic ray electron spectra \citep{wiebel1999}.  So, at radio frequencies \mbox{$\nu > \nu_2$} the radio
emission is again weaker, this time by $(\nu/\nu_2)^{-1/3}$.  So again
the radio emission is slightly down. Both variants show, that the radio
emission is not far from the diffusive equilibrium.

This concludes the demonstration of the argument. There are many checks on
these ideas, which one can make, such as the pressure of the interstellar
medium, the X-ray luminosity, the possibility to account for extreme
galaxies such as M82, the radial gradient of the far-infrared/radio
ratio \citep{bicay1989}, the thickness of the hot gaseous disk and the 
associated
diffusion coefficient of cosmic rays, and many others. We will discuss these
points elsewhere.

\subsection{Where is the limit to the diffusion limit?}
The line of reasoning seems to be a deus ex machina in the sense that we 
seem to be always in the diffusion limit, independent of whether we have 
starburst galaxies with relatively strong magnetic fields, or normal 
galaxies like our own.

The equations above imply that
\begin{equation}
\tau_{diff} \; = \; \frac{(2 k_B T)^{2}}{\Lambda} \frac{8 \pi}{B^{2}} \; 
< \; \frac{6 \pi m_e c}{\sigma_T \gamma_e B^{2}} \; = \; \tau_{syn}\,.
\end{equation}
We note first that both sides depend on the magnetic field strength 
squared, so the argument is independent on the magnetic field: Inserting 
numbers we find

\begin{equation}
\frac{(T_5)^{2}}{\Lambda_{-21}} \; < \; \frac{4 \cdot 10^{7}}{\gamma_e}\,,
\end{equation}
where $\gamma_e$ is the Lorentz factor of the mainly emitting
electrons.  This time, however, we need to ask what density contrast the
main radio emitting substructures have, since the left hand side of the
equation uses grand total volume averages, and the right hand samples
just those regions, where $B^{2}$ is especially high.  Already simple
arguments, as shown above, demonstrate that the average of $B^{2}$  is
quite a bit higher than the average of $B$ squared.  We can assume here,
that such a factor might be of order 30, or even higher, bringing the
limiting energy of the electrons down to \mbox{$\gamma_e < 10^{6}$}, or
possibly even much less.
We have already argued earlier that for the cooling to be maximal, the
temperature has to be close to or below $10^{5}$ K.

This allows us to understand perhaps, why galaxies are usually in the 
diffusion limit, sometimes of course, in a starburst, in the injection 
limit: In that case the overall spectrum corresponds to injection. There 
was just not enough time to achieve diffusive approximate equilibrium.
\subsection{Implications}
First of all, this theory does just what one would naively expect,
relate the
massive stars which heat the dust through their ultraviolet light directly
with the subsequent supernovae. All the theory does, is work out the
non-linearity inherent in synchrotron emission, and shows them to introduce
negligible dependencies on various parameters.

Starburst galaxies as well as quiescent galaxies equally obey the 
correlation
between the radio and the far-infrared emission. Therefore the theory
implies
by necessity that the magnetic field in a starburst region rises with
the
overall energy density.

Galaxies which are subject to substantial compression by an encounter with
another galaxy, clearly have a magnetic field which is higher than 
corresponds
to the energy density derived from star formation, and so may be
expected to
have more radio emission than indicated by the general radio-far-infrared
correlation. This is borne out by at least one example \citep{hummel_beck1995}.

Furthermore, it is clear that for very short time scales, near order $10^7$
years, the correlation cannot hold, since we then run into the
lifetimes of
the massive stars, which drive the energy balance in the interstellar 
medium.

There is likely also a lowest level of star formation activity, where the
assumption that the hot medium is fully connected, fails. There one may 
expect
also substantial departures from the correlation.

\section{Cosmic rays and their secondaries\label{cosmic_rays:sec}}
As stated above, the cosmic ray intensity from starburst galaxies
scales with the radio and infrared emission of the
sources. In this section, we discuss the emission scenarios for charged cosmic
rays and hadronic interactions leading to high-energy photon and
neutrino emission. 
There are two source classes within starbursts that can accelerate cosmic rays to high energies,
namely shock fronts of supernova remnants and long Gamma Ray Bursts
(GRBs), the latter being
connected to supernova Ic explosions. In the first case, maximum energies are
limited to less than $10^{15}$~eV and thus, the cosmic rays from starbursts cannot be
observed directly due to the high cosmic ray background in our own
Galaxy. Gamma Ray Bursts, on the other hand, were proposed as the origin of
cosmic rays above the ankle, i.e.\ \mbox{$E_{\rm CR}>3\cdot 10^{18}$ eV},
see \cite{vietri1995,waxman1995}. Since a high star formation rate as
it is present in starburst galaxies, leads to a high rate supernova
explosions, an enhanced rate of long GRBs is expected. Thus, for
closeby sources, the distribution of starburst galaxies can be used to test
the hypothesis of cosmic rays from starbursts, as also discussed
in \cite{biermann_review2008}.

The dominant source of secondary cosmic rays like high-energy photons and
neutrinos is proton-proton interactions in dense hydrogen regions and proton-photon interactions in Gamma Ray Bursts.
Proton-proton interactions produce pions via 
\begin{equation}
p\,p\rightarrow \pi^{+}\,\pi^{-}\,\pi^{0}\,.
\label{pp:equ}
\end{equation}
Photohadronic interactions, on the other hand, yield pions via the Delta
resonance,
\begin{equation}
p\,\gamma\rightarrow \Delta^{+}\rightarrow
n\,\pi^{+}/p\,\pi^{0}\,.
\end{equation}
High-energy photons and neutrinos are subsequently emitted in
$\pi^{\pm}-$ resp.\ $\pi^{0}-$decays:
\begin{eqnarray}
\pi^{+}&\rightarrow&\mu^{+}\,\nu_{\mu}\rightarrow
e^{+}\,\nu_{e}\,\overline{\nu}_{\mu}\,\nu_{\mu}\nonumber\\
\pi^{-}&\rightarrow&\mu^{-}\,\overline{\nu}_{\mu}\rightarrow
e^{-}\,\overline{\nu}_{e}\,\nu_{\mu}\,\overline{\nu}_{\mu}\nonumber\\
\pi^{0}&\rightarrow&\gamma\,\gamma\,.\label{pion_decays:equ}
\end{eqnarray}
\subsection{Supernova remnants}
Cosmic Rays are believed to be produced in young supernova remnants (SNRs),
reaching maximum energies of around $10^{15}$~eV or above, depending
on their local environment and on the cosmic ray composition. The production of
secondaries from hadronic interactions depends on the proton-proton
optical depth in the SNR environment. In this section, we present a
model of which sources are optically thin and which, in contrast, are good
candidates for the production of high-energy photons and neutrinos. 

\subsubsection{Optical depth\label{opt_depth:sec}}
\begin{figure}
\centering{
\epsfig{file=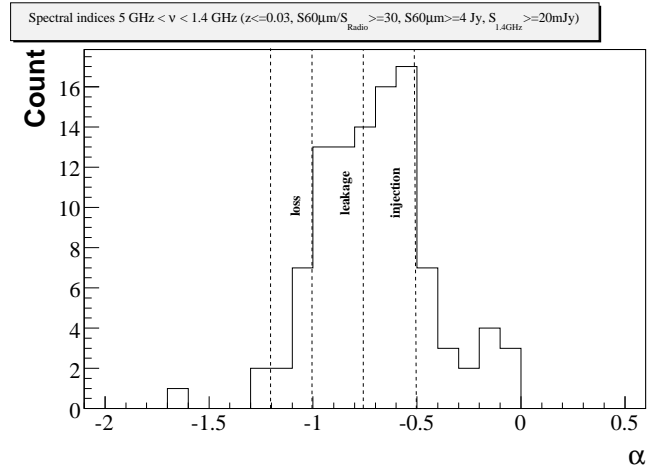,width=\linewidth}
\caption{Histogram of the radio spectral indices of 105 sources between $1.4$~GHz and
  $5$~GHz. The areas between the dashed lines indicate sources in the
  loss, leakage and injection limit (from the left). As a
  conservative estimate, we include those three sources with
  extremely steep spectra as loss limit sources. Those sources with
  extremely flat spectra, to the right of the injection limit area,
  are dominated by absorption or free-free radiation. \label{spectral_indices_14_5}} 
}
\end{figure}

We use the observation of synchrotron radiation from shock-accelerated
electrons, assuming that electrons and hadrons are accelerated in the
same shock environment. Figure~\ref{spectral_indices_14_5} shows the distribution of spectral indices at radio wavelengths
between $\oneghz$ and $\fiveghz$ for 105 sources in the sample with given
spectral indices at the required wavelengths. The spectrum at these energies
is produced by electron synchrotron losses. Depending on the shape of the
primary electron spectrum and scattering effects, the spectral index of the
electron population can reach different
values. Shock acceleration of charged particles usually results in primary electron
spectra of
\begin{equation}
\diffe\propto {\eel}^{-\alphain}\,,
\end{equation}
with \mbox{$\alphain\approx2.0-2.4$}. If the electrons escape before interacting with
the ambient medium, the primary spectrum stays unmodified, \mbox{$\alphain=\alphaem$}, referred to as the {\it
  injection limit}. If the electrons are partly
scattered down to lower energies, the spectrum of secondaries steepens to $\alphaem\approx
2.5-2.8$ for a primary spectrum with $\alphain\approx 2.0-2.4$. This is called the {\it leakage limit}. In the case of calorimetric sources,
basically the entire energy is lost in the source and the spectrum of secondary electrons is as steep as $\alphaem\approx 3.2-3.4$. This scenario is
referred to as
the {\it loss limit}. 

As discussed in~\cite{rybicki1979}, the electron spectral index
$\alphaem$ correlates with the index of synchrotron radiation
as
\begin{equation}
\alpha=-\frac{\alphaem-1}{2}\,.
\end{equation}
Thus, the observed synchrotron spectral indices are
\begin{equation}
\alpha=\left\{
\begin{array}{lll}
-0.5\rightarrow -0.75&\rm in\, the&\rm {\it injection\, limit}\\
-0.75\rightarrow -1.0&\rm in\, the&\rm {\it leakage\, limit}\\
-1.0\rightarrow -1.2&\rm in\, the&\rm {\it loss\, limit}\,.\\
\end{array}
\right.
\end{equation}
The synchrotron spectral indices in this sample scatter between
\mbox{$-1.7<\alpha<0.7$} with a peak at \mbox{$\alpha \sim -0.7$}. We exclude sources that
may include contributions other than synchrotron radiation, i.e.~those sources
that have spectral indices \mbox{$\alpha>-0.5$}. Here, absorption is likely
to have modified the spectrum. Alternatively, spectral indices around
-0.1 point to free-free radiation \citep{mezger_henderson1967}. Of the
remaining 85 sources, 37 sources (44\%) starbursts are in the
injection limit, 36 sources (42\%) are in the leakage limit and
12 sources
(14\%) are in the loss limit.

The observation of a synchrotron spectrum following the injected electron
spectrum has severe implications for hadrons in the same source. As proton
interactions require much higher particle or electromagnetic field densities
than electron interactions, it can be expected that hadrons are not affected
if electrons can escape freely. Thus, protons escape from the source before
they interact. Even in the leakage limit, it is unlikely to have proton
interactions with matter or photons, only sources in the loss limit provide
conditions with reasonable densities for proton interactions. In  the case of
high particle densities, protons interact with each other and produce pions as
described in Equations (\ref{pp:equ}) and (\ref{pion_decays:equ}).
Hence, only those 14\% of all starburst sources in the loss limit are likely to produce both high-energy
neutrinos and photons. We will take this into account in all following
calculations by applying a factor of \mbox{$\epsilon_{loss}=0.14$}, so that
only loss limit sources are included.
\subsubsection{Contributions of starbursts to the FIR background\label{fir_bg:sec}}
In a previous estimate of high-energy neutrino radiation
from starburst galaxies by~\citep{lw2006}, it is assumed
that the entire background of far infra-red radiation comes from
starburst galaxies. However, it is pointed out by~\cite{stecker_sbg06}
that only a fraction of 23\%
the total diffuse FIR background actually originates from
starbursts. This would reduce the flux of hadronic secondaries by a
factor \mbox{$\xi_{FIR}=0.23$}.
Still, as it is pointed out by~\cite{thompson2006}, a fraction near unity,
\mbox{$\xi_{FIR}\approx 1$}, is consistent with most star formation rate models today and
should be considered as an upper limit estimate. 
In this paper, we
will use the FIR background from the EBL model given by
\citep{kneiske2002}, where a total of 80\% comes from starbursts.
\subsubsection{Production of hadronic secondaries\label{production:sec}}
The non-thermal radio emission from starburst galaxies indicates the shock
acceleration of electrons. Hadrons are accelerated in the same way. In the
case of proton interactions, high-energy neutrinos and photons can be produced as indicated in
Equations (\ref{pp:equ}) and (\ref{pion_decays:equ}). For heavier nuclei, the efficiency is slightly less than
for protons due to photo-disintegration, see
e.g.~\cite{hts2005,ave2005,anchordoqui2008}. In this paper, we calculate
secondary spectra from protons, and consider heavier cosmic rays elsewhere. High-energy photon emission can also be due to
bremsstrahlung or inverse Compton scattering, but above $1$~GeV, pion decay
photons should dominate as discussed in~\cite{paglione1996,domingosantamaria_torres2005}.
Here, we re-calculate the possible neutrino and
photon fluxes for those sources optically thick to
proton-proton interactions. Normalization, spectral behavior and
propagation  are treated
as follows:
\noindent
\begin{itemize}
\item \emph{Spectral behavior}\\
We assume that the protons at the source follow a power-law spectrum with
an index $\alpha_p$ and with exponential
cutoff at $E_{\max}$,
\begin{equation}
\frac{dN_p}{d\ep}=A_p\cdot {\ep}^{-\alpha_p}\cdot
\exp\left[-\frac{\ep}{E_{\max}}\right]\,.
\label{proton_spect:equ}
\end{equation}
The normalization of the spectrum $A_p$ is determined by assuming that
a fraction of the total SNR energy, $\eta$, goes into cosmic
rays. This is described in more detail in the next paragraph. The
observed spectrum of cosmic rays below the knee, i.e.\ below $10^{15}$~eV is typically
assumed to come from supernova explosions in the Galaxy, with a cutoff
at $10^{15}$~eV. The spectral part between the knee and the ankle of
the cosmic ray spectrum is still a matter of debate, but may arise
from the heavy particle component from SNRs, see \citep{stanev1993}. Here, we assume that SNRs in starburst galaxies
produce similar spectra. Protons are likely to have an energy cutoff
even below $10^{15}$~eV. As the exact cutoff energy is not known, we
use \mbox{$E_{\max}=10^{15}$ eV} as an upper limit. The observed
spectral index is $\alpha_p=2.7$. Stochastic particle acceleration
usually produces spectra of around \mbox{$2.0<\alpha_p<2.4$}. Hence, it is
not sure yet where the steepening of the spectrum occurs, whether it
is an internal steepening or a propagation effect. We therefore test
proton spectra with indices of \mbox{$\alpha_p=2.7,\,2.4,\,2.2,\,2.0$}. A
spectrum of $E^{-2.7}$ would be present when diffusion applies,
spectral indices between $2.0$ and $2.4$ are predicted
by stochastic acceleration without significant diffusion. This approach differs from the
calculations by \cite{paglione1996,domingosantamaria_torres2005}, who use the diffusion-loss equation to
determine the spectral index.\\[0.2cm]
In all following calculations, we determine the spectra of hadronic
secondaries from proton-proton interactions, using the delta-functional
approximation for proton energies \mbox{$E_p<100$ GeV}, see e.g.~\citep{schlickeiser_mannheim1994}. At higher energies,
the more exact
analytic approximation as presented
in~\cite{kelner2006} is used, where Monte-Carlo simulation results are
approximated by analytical equations. For all energies, the
logarithmic increase of the proton-proton cross section with energy is taken into account as
described in~\cite{kelner2006}. We have a high hydrogen density,
\mbox{$n_H=100$ cm$^{-3}$}, as we expect the dominant proton acceleration
and interaction to occur in heavy supernova remnants, having red
supergiants or Wolf-Rayet stars as progenitors. 
\item \emph{Normalization}\\
We normalize the energy spectrum of cosmic rays by determining their
energy density $\rho_{CR}$.
The latter is defined by the energy
  integration over the differential cosmic ray flux, multiplied
  with the energy, multiplied by a factor \mbox{$4\pi/c$} to get from a flux
  to a density,
\be
\rho_{CR}:=\frac{4\pi}{c}\int_{E_{\min}}^{\infty}dE_p\,\frac{dN_p}{dE_p}\cdot
E_p\,.
\ee
In order to determine the normalization of the total proton flux from
starbursts, we assume a power-law spectrum with an exponential cutoff
as described in Equ.~(\ref{proton_spect:equ}).\\[0.2cm]
The energy density, in turn, is directly proportional to the supernova rate in a
  galaxy, $\dot{n}_{\rm SN}$, assuming that a fraction $\eta$ of the total energy of a supernova,
  \mbox{$E_{\rm SNR}\approx 10^{51}$ erg} is transferred to cosmic
  rays,
\begin{equation}
\rho_{CR}=\frac{\epsilon_{loss}\cdot\eta\cdot E_{\rm SNR}\cdot \dot{n}_{\rm SN}}{c\cdot
  d_{l}^{2}(z)\cdot (1+z)^2}\,.
\end{equation}
Here, we use \mbox{$\epsilon_{loss}=0.14$} as discussed
in Sections~\ref{opt_depth:sec} and \ref{fir_bg:sec}. We further assume that 5\% of the total
SNR energy is transferred to cosmic rays, i.e.~$\eta=0.05$.
To determine the supernova rate in a galaxy, the total FIR luminosity at a
given redshift $z$, \mbox{$\lfir^{tot}(z)$}, is estimated from the FIR emissivity
\mbox{$\mathcal{E}_{FIR}(z)$}, the latter coming from the extragalactic
background light (EBL) model from~\cite{kneiske2002},
 \begin{equation}
{\lfir}^{tot} = \int \mathcal{E}^{FIR}_{\nu}(z)\ \frac{dV}{dz}(z)\mathrm d\nu\,.
\end{equation}
Here, $dV/dz$ is the comoving volume element.\\[0.2cm]
The supernova rate was determined
by~\cite{mannucci2003} to correlate with the FIR luminosity of
the galaxy,
\begin{equation}
\dot{n}_{\rm SN}=(2.4\pm0.1)\cdot 10^{-12}\cdot
\left(\frac{\lfir}{L_{\odot}}\right)\,{\rm yr}^{-1}\,.
\label{snr_rate:fig}
\end{equation}
The FIR luminosity is expressed in terms of the solar luminosity,\\
\mbox{$L_{\odot}=3.839\cdot 10^{33}$}~erg/s. This relation was 
predicted in~\citep{biermann_fricke1977,kronberg_biermann1981} within
a factor of 3, and other experimental results
from \cite{vanburen_greenhouse1994} yield the same correlation within
uncertainties.
Finally, the cosmic ray energy density is given as
\begin{eqnarray}
\rho_{CR}&=&1.6\cdot
10^{-13}\,\mbox{erg/cm}^{3}\cdot\nonumber\\
&\cdot&\epsilon_{loss}\cdot 
\frac{\eta}{0.05}\cdot \frac{E_{SNR}}{10^{51}\,{\rm erg}}\cdot
\frac{L_{FIR}}{10^{12}\,L_{\odot}}\nonumber\\
&\cdot&
\left(\frac{d_{l}(z)}{6.5\,{\rm Gpc}}\right)^{-2}\cdot
  \left(\frac{1+z}{2}\right)^{-2}\,.
\end{eqnarray}
This number is compatible with the observed cosmic ray spectrum above
$1$~GeV in the Galaxy, \mbox{$\rho_{CR}^{MW}=1$ eV/cm$^{3}$}, assuming a supernova
rate of \mbox{$\dot{n}_{SN}=0.03/$yr}.
\item \emph{Propagation}\\
We apply that the particle energy at Earth is a factor of $1+z$ higher
than at the source. Neutrinos travel in straight lines without
interaction. In the case of photon propagation, we include absorption
effects by the EBL, using the model of \cite{kneiske2002}.
\end{itemize}
From the considerations above, the spectra of hadronic secondaries at Earth from a given
redshift $z$ are determined and a simple redshift integration is
performed in order to get the total neutrino flux at Earth. Here, we
integrate from $z=0$ up to a redshift where first starbursts are
expected to be formed. We use \mbox{$z_{\max}=5$}, since the main contribution to the
spectra comes from redshifts up to \mbox{$z\sim1-2$}. The contribution above
\mbox{$z=5$} is negligible.
\begin{table}
\centering{
\begin{tabular}{l|l|l}
\hline
quantity&variable&value\\\hline\hline
total energy release (SNR)&$E_{SNR}$&$10^{51}$~erg\\
energy fraction transferred from SNR to CRs&$\eta$&$0.05$\\
hydrogen density&$n_H$&$100$~cm$^{-3}$\\
fraction of starbursts in the loss limit&$\epsilon_{loss}$&0.14\\
\end{tabular}
\caption{Parameters used to determine the diffuse flux of high-energy photons
  and neutrinos from SNRs in starburst galaxies.\label{params_crs:tab}}
}
\end{table}
\subsubsection{High-energy photons}
Assuming proton-proton interactions in starbursts as described above,
we calculate the gamma-ray emission produced by starburst galaxies
for energies \mbox{$>$ 1 GeV}. A detailed calculation and discussion for the starburst
galaxy NGC253 has been done in \cite{paglione1996,domingosantamaria_torres2005}. 
Gamma-ray emission in starburst galaxies is due to three different processes, 
Bremsstrahlung, inverse Compton scattering and pion decay. In the
following only gamma-ray emission from pion decay is considered
because it is the dominating process at energies above 1~GeV
\citep{paglione1996,domingosantamaria_torres2005}. 

\begin{figure*}
\centering{
\epsfig{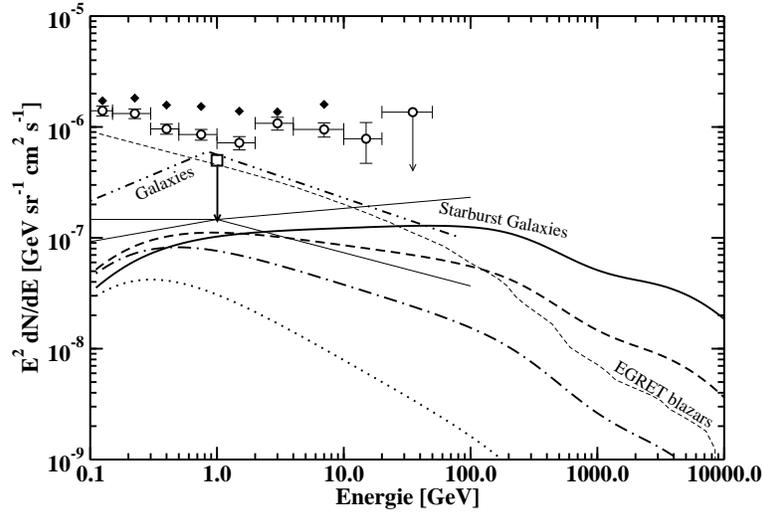}
\caption{Expected diffuse high-energy photon flux from SNRs in
starburst galaxies. The black lines represent the contribution for an
$E^{-2}$ (thick, solid), $E^{-2.2}$ (thick, dashed), $E^{-2.4}$
(thick, dot-dashed) and $E^{-2.7}$
(thick, dotted) proton input
spectrum with an exponential cutoff at $10^{15}$~eV. Closed data points are
from the EGRET experiment~\citep{egret}. Open data points represent an
update of the EGRET data, using an updated model for galactic
gamma-ray emission~\citep{egret_new}. The thin, dashed
line represents the contribution from EGRET
blazars~\citep{kneiske_mannheim2007}. The thin, dot-dot dashed
line shows the possible contribution from regular galaxies \citep{pavlidou_fields2002},
while the thin, solid lines 
display
contribution from starbursts as calculated
by~\cite{thompson2007}, $E^{-2}$ and $E^{-2.3}$ initial proton spectra. 
\label{photons:fig}}} 
\end{figure*}

In Fig.~\ref{photons:fig}, the model of gamma-ray emission from starbursts as
presented above is shown for three different proton spectra,
i.e.~$E^{-2.0}$ as the thick solid, $E^{-2.2}$ as the thick dashed, $E^{-2.4}$
as the thick dot-dashed and $E^{-2.7}$ the thick dotted line. We use that only 14\% of all starbursts are
in the loss limit \mbox{($\epsilon_{loss}=0.14$)} and we apply the FIR
background from the EBL model by \cite{kneiske2002}. Absorption due to
the EBL is considered in all calculations. We compare our results to the observation of the
diffuse, extragalactic background as measured by
EGRET~\citep{egret}. Closed squares represent the first analysis
presented in~\cite{egret}, while open circles show an updated
analysis, using an improved model for the distraction of the galactic
component from the extragalactic contribution~\citep{egret_new}. The
high-energy radiation from starburst galaxies makes up about 10\% percent
of the total background. In the same figure, we
show other possible contributions to the total diffuse background,
like the one from resolved EGRET blazars \citep{kneiske_mannheim2007}
or the contribution from regular
galaxies~\citep{pavlidou_fields2002}. \cite{karlsson2008} present a model
of high-energy photon emission from starbursts. Just as
\cite{karlsson2008}, we include the logarithmic rise of the proton-proton
cross section with energy and get comparable results. In addition, our
results yield a slightly lower flux than the prediction by \cite{thompson2007}. 
The main reason is that we include that only 14\% of all
starbursts are proton calorimeters where \cite{thompson2007} assume
that this fraction increases from 10\% for local starbursts to a
saturated value of 80\% at \mbox{$z=1$}. We use this
more conservative approach.
The considerations above show that several factors are quite uncertain,
i.e.~the total energy released by a supernova remnant, the hydrogen
density, as well as the number
of starburst galaxies that contribute to the FIR background and that
are proton calorimeters. Thus, the possibility of
starbursts contributing significantly to the diffuse photon background
at high energies should be considered. The Fermi
satellite\footnote{called GLAST before its launch}, launched
on June 11, 2008, will help identifying starburst galaxies at energies above
$100$~MeV, to determine the exact fraction of the starburst diffuse
photon flux in the total gamma background. 

\subsubsection{High-energy neutrinos}
\begin{figure*}
\centering{
\epsfig{file=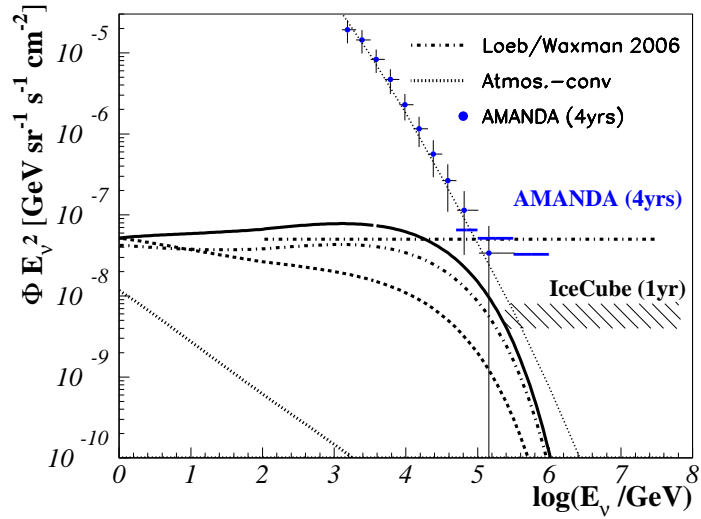,width=10cm}
\caption{Expected diffuse high-energy neutrino flux from SNRs in
starburst galaxies. The black lines represent the contribution for an
$E^{-2}$ (thick, solid), $E^{-2.2}$ (thick, dashed), $E^{-2.4}$
(thick, dot-dashed) and $E^{-2.7}$ (thick, dotted) proton input
spectrum with an exponential cutoff at $10^{15}$~eV. Data points show
the atmospheric neutrino background as measured by the AMANDA
experiment (data between 2000 and 2003) \cite{kirsten_icrc2007,kirsten_phd}. The
prediction of atmospheric neutrinos is taken from~\cite{volkova1980}. AMANDA
limits are for the same data sample, derived from the fact that no
significant excess above the atmospheric background was observed. The
dot-dashed line shows the prediction by \cite{lw2006}, not taking
into account that only 14\% of all starbursts are calorimeters. \label{starburst_diffuse:fig}} 
}
\end{figure*}

The high-energy neutrino flux calculated according to Section~\ref{production:sec}
is presented in Fig.~\ref{starburst_diffuse:fig}, for an
$E^{-2.0},\,E^{-2.2},\,E^{-2.4},\,E^{-2.7}$ (thick solid, dashed,
dot-dashed and dotted lines) spectral behavior of the initial
proton spectrum. Although, in contrast to \cite{lw2006}, we take into account that only 14\% of all
sources are effective proton calorimeters, the flux strength of our prediction is
compatible with \cite{lw2006}: While \cite{lw2006} use the radio
luminosity of local starbursts as an estimate, we take the supernova
rate as a measure. The
reason for the cutoff at lower energies is that we
assume that protons are not accelerated beyond $10^{15}$~eV, which is
based on the observation of the knee in cosmic rays at about
$10^{15}$~eV. The interpretation that SNRs in the Galaxy are responsible for the
flux below $10^{15}$~eV justifies the assumption that also in
starbursts, SNR accelerate cosmic rays up to similar energies. It
should be kept in mind, however, that the actual cutoff for protons is
likely to lie at energies below $10^{15}$~eV and that only heavier
cosmic rays, which we do not consider in this calculation, can reach
energies as high as $10^{15}$~eV.
With an energy cutoff at or below $10^{15}$~eV, the neutrino flux from
starbursts is out of reach for the detection of a diffuse flux by high-energy neutrino
detectors like Km3NeT and IceCube as indicated in
Fig.~\ref{starburst_diffuse:fig}. 

\subsection{Gamma Ray Bursts}
As starburst galaxies show an enhanced rate of
supernova explosions, an increased rate of long Gamma Ray Bursts (GRBs), which
are directly linked to SN-Ic events~\citep[e.g.]{grb030329}, is expected. Thus, if long GRBs are the dominant sources of
UHECRs, the contribution from nearby objects should follow the distribution of starburst
galaxies. Here, we examine the number of GRBs to be expected from our
catalog.

\begin{figure}
\centering{
\epsfig{file=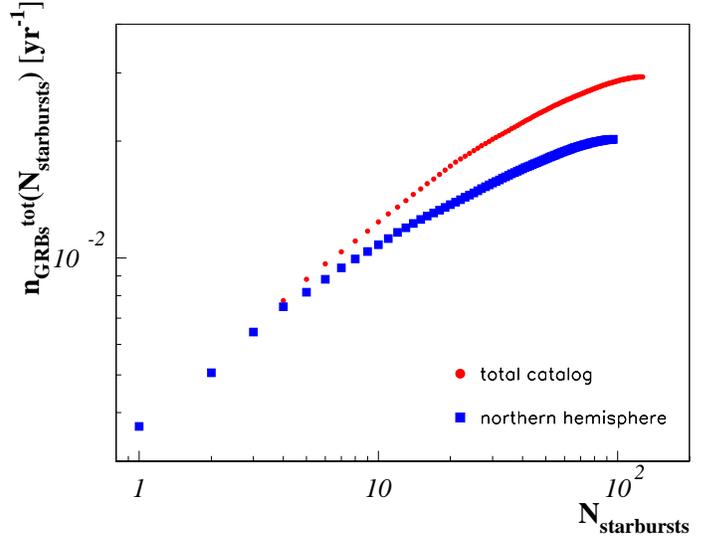,width=10cm}
\caption{Number of GRBs per year in the starburst catalog, including
  $N_{starbursts}$ sources, starting with the strongest one. Red circles
  include all starbursts in the sample. The total GRB rate in the
  sample, including all 127 sources, is
  \mbox{0.03~yr$^{-1}$}, which means that a GRB could be observed every 30 years on
  average. Results shown in blue squares only
  show northern hemisphere bursts, lying in the IceCube FoV. Northern
  hemisphere GRBs would occur every 50 years \mbox{($\dot{n}_{GRB}= 0.02$~yr$^{-1}$)}.\label{grbs_year:fig}} 
}
\end{figure}

In the following calculations, we assume that every SN-Ic explosion is accompanied by a particle
jet along the former star's rotation axis, i.e.~by a GRB. The opening angle of
the GRB jet $\theta$ determines, how many SN-Ic can be observed as GRBs,

\begin{equation}
\dot{n}_{\rm GRB}=\epsilon\cdot \dot{n}_{\rm SN-Ic}\,.
\label{grbs_snic:equ}
\end{equation}
Here, $\dot{n}_{\rm GRB}$ is the GRB rate in a
galaxy and \mbox{$\epsilon=(1-\cos\theta)$} is the fraction of SN-Ic producing
GRBs. The jet opening angle is difficult to determine. Typically, one expects
an opening angle of less than $10^{\circ}$ for the prompt emission, see e.~g.~\cite{berger2003,grb080319b_nature_konus}. Afterglow
emission and precursors can have larger opening angles~\citep{morsony2007}. As we focus on the
prompt emission, we will use a typical opening angle
 of $\sim 10^{\circ}$ as an optimistic estimate, i.e.
\begin{equation}
\epsilon \approx 0.015\,.
\end{equation}
Further, observational data show that core collapse supernovae of type SN-Ib/c
contribute with 11\% to the total SN rate in starbursts~\citep{cappellaro_turatto2001}.
Thus, using Equ.~(\ref{grbs_snic:equ}), the GRB rate in a
starburst galaxy is
directly correlated to the supernova rate $\dot{n}_{\rm SN}$,
\begin{equation}
\dot{n}_{\rm GRB}=\epsilon\cdot \zeta \cdot \dot{n}_{\rm SN}\,,
\label{grbs_sn:equ1}
\end{equation}
with \mbox{$\zeta\sim 0.11$} as the fraction of heavy SN explosions in all SN
explosions in a single galaxy. Using Equ.~(\ref{snr_rate:fig}) to determine the supernova
rate in a galaxy in
Equ.~(\ref{grbs_sn:equ1}) yields a GRB rate of
\begin{equation}
\dot{n}_{\rm GRB}=3.8\cdot 10^{-15}\cdot
\left(\frac{\lfir}{L_{\odot}}\right)\cdot
  \left(\frac{\epsilon}{0.015}\right)\cdot\left(\frac{\zeta}{0.11}\right) \,{\rm yr}^{-1}
\end{equation}
per starburst.
With an expected lifetime of more than $10$~years for a neutrino detector like
IceCube, luminosities of around \mbox{$3\cdot 10^{13}\cdot
L_{\odot}\sim 10^{47}$}~erg/s are required for the detection of
a single event. None of the sources in our catalog provides such high
luminosities: {\sc IRAS17208-0014} is the intrinsically strongest
source with \mbox{$\lfir=5.9\cdot 10^{45}$}~erg/s, the second
strongest one is {\sc IRASF17207-0014} with \mbox{$\lfir=5.5\cdot
10^{45}$}~erg/s. If, however, a
larger number of starbursts is considered for an analysis, the total
luminosity increases and with it the probability of observing a
GRB. Figure~\ref{grbs_year:fig} shows the total GRB rate for a number
of $N_{\rm starbursts}$ galaxies,
\be
\dot{n}_{\rm GRB}^{\rm tot}(N_{\rm starbursts})=\sum_{i=1}^{N_{\rm
    starbursts}}\dot{n}_{\rm GRB}(i{\rm th\,\, starburst})\,.
\ee
In the figure, we sum up the
GRB rates achieved in the single starbursts, starting with the most
luminous source, adding sources in descending luminosity order.  The
red points show the GRB rate summing over all starbursts in the
sample. On total, $0.03$~GRBs per year are expected to be observable
from our sample of starbursts. The blue squares display the total GRB rate, summing up
sources in the northern hemisphere, which corresponds to IceCube's
Field of View (FoV), in order to estimate the neutrino detection probability in
Section~\ref{neutrinos_grbs:sec}. Here, $0.02$~GRBs per year are
expected. This number can be enhanced significantly when taking into
account those weaker sources which we do not include in our catalog in
order to ensure completeness (see Section \ref{catalog:sec}). 
\subsubsection{Observation of cosmic rays and starburst galaxies}
Of the 127 sources in our catalog, 96 are located in the northern
hemisphere. Thus, Auger South will only have very few starbursts in its FoV,
while HiRes, the Telescope Array (TA) and later Auger North will be able to observe a large fraction of
starbursts. As already discussed by~\cite{biermann_review2008}, the question
of the origin of UHECRs can only be resolved by taking this strongly
asymmetric distribution into account. Even if GRBs contribute to the total
flux of cosmic rays, nearby sources cannot be identified by a southern
hemisphere telescope, since the number of sources is too small. Given the
large number of sources in the northern hemisphere, telescopes like HiRes, TA and Auger North can investigate this matter. In
particular, for GRBs coming from one of the starbursts in our catalog could
give the opportunity of an enhanced UHECR flux within a short time window,
assuming that the signal is still focused in time and space due to the
closeness of the sources. Thus, an analysis with UHECR detectors in the
northern hemisphere for nearby GRBs could be optimized by not only looking for
spatial but also for temporal clustering.
\subsubsection{Enhanced neutrino flux from GRBs in starbursts\label{neutrinos_grbs:sec}}
 While the detection of a
permanent, diffuse signal from GRBs in nearby starbursts will not be possible due to the high
atmospheric background seen by high-energy neutrino telescopes, a timing analysis may be able to
identify the Gamma Ray Bursts in neutrinos. In such an analysis, the location of a
nearby starburst can be chosen as a potential neutrino hot-spot. By
defining a time window of the typical duration of a long GRB \mbox{($\sim
100$~s)}, the atmospheric background can be reduced to close to
zero. The GRB event rate per starburst galaxy was already examined
above. Here, the general neutrino intensity and in particular the possibility
of detection with IceCube are
  discussed.
\paragraph{Neutrino spectrum from a single GRB in a starburst}
The neutrino spectrum during the prompt photon emission
phase in a GRB was for the first time determined
by~\cite{wb1997,wb1999}. It can be expressed as a broken power-law,
\begin{equation}
\frac{dN_{\nu}}{d\en}=A_{\nu}\cdot {\en}^{-2}\cdot \left\{ \begin{array}{lll}
{\en}^{-\alpha_{\nu}+2}\cdot{\enb}^{\alpha_{\nu}-\beta_{\nu}} &&\mbox{ for } \en<\enb\\
\en^{-\beta_{\nu}+2}&&\mbox{ for } \enb<\en\leq \ens\\
\ens \cdot \en^{-\beta_{\nu}+1}&&\mbox{ for }\en> \ens\,.
\end{array}
\right.
\end{equation}
The spectrum includes the five parameters
$\alpha_{\nu},\,\beta_{\nu},\,\enb,\,\ens,\,A_{\nu}$. We discuss the numerical
values of the parameters below. For their derivation, see
e.g.~\cite{guetta2004,bshr2006}. The derivation is done for an isotropically
emitted signal. However, the neutrino spectrum itself does not vary with the
jet's opening angle, since the parameters are determined by energy
densities. Where the energy is enhanced by a factor \mbox{$1/(1-\cos\theta)$}, the
solid angle decreases just as \mbox{$(1-\cos\theta)$} and the factors cancel.
\begin{itemize}
\item {\it The break energies $\enb$ and $\ens$}\\
The neutrino spectrum can be derived when assuming that the protons
accelerated along the GRB jet interact with the ambient synchrotron
photon field. Neutrinos are produced in $\pi^{+}-$decays as
described in Equ.~(\ref{pion_decays:equ}). The first break energy
represents the energy required to produce the Delta resonance in
proton-photon scattering. At higher energies, $\en>\enb$, the
neutrino spectrum follows the proton spectral behavior,
\mbox{$\beta_{\nu}\sim \alpha_p$}. At lower energies, \mbox{$\en<\enb$}, scattering becomes less
effective and the spectrum becomes flatter, \mbox{$\alpha_{\nu}\sim
\alpha_p-1$}. Due to the transformation from the CM system of the
proton-photon interaction in the source into the observer's frame at Earth, the first break energy depends
on the shock's boost factor \mbox{$\Gamma:=10^{2.5}\cdot \g25$}, the observed
photon break energy of GRBs and
the redshift (\mbox{$z\approx 0$} for our catalog),
\begin{equation}
\enb=7\cdot10^{5}\cdot (1+z)^{-2}\,\frac{\g25^{2}}{\egbM}\mbox{
  GeV}\approx 3\cdot 10^{6}\mbox{
  GeV}\,.
\label{enb:equ}
\end{equation}
We fix the boost factor to \mbox{$\g25=1$} and the observed photon break
energy to \mbox{$\egbM:=\egb/$MeV$\sim 0.25$}. For the sources
in our catalog, we have \mbox{$z\approx$}~$0$.

The second break in the neutrino spectrum, due to
pion-synchrotron losses, is determined by unknown parameters like
variability time scale \mbox{$t_{v}\approx 0.01$~s~$\cdot t_{v,-2}$}, GRB luminosity \mbox{$L_{\gamma}\sim 10^{51}$~erg/s~$\cdot \lumi$} and electron and magnetic field equipartition
fractions, \mbox{$\epsilon_e\approx \epsilon_B\approx$}~$0.1$. 
\begin{equation}
\ens=\frac{3\cdot 10^{7}}{1+z} \epsilon_{e}^{1/2} \, \epsilon_{b}^{-1/2}\, \g25^{4}
\,t_{v,-2}/\sqrt{\lumi} \mbox{ GeV}\approx 3\cdot 10^{7} \mbox{ GeV}\,.
\label{ens:equ}
\end{equation}
Here, we fix the parameters to\mbox{ $t_{v,-2}=1$}, \mbox{$\lumi=1$}, \mbox{$\epsilon_{b}=\epsilon_{e}=$}~$0.1$.
\item {\it The spectral indices $\alpha_{\nu}$ and $\beta_{\nu}$.}\\
In the energy range \mbox{$\enb<\en<\ens$}, the spectral index
\mbox{$\beta_{\nu}=p$} is represented by the primary hadron spectral index. 
We assume a simple $E^{-2}$ proton spectrum in this case: While it was argued
previously, that highly relativistic flows have a limit of not
becoming flatter than $\sim E^{-2.2}$ \citep{bednarz_ostrowski98}, recent studies
of oblique shocks have shown that particle spectra as flat as
$E^{-1.5}$ can be produced with shock boost factors of \mbox{$\Gamma>$}~$100$ as
they occur in GRBs, see~\cite{mbq2008}. In addition, including large angle
scattering of particles in the acceleration process also leads to flat particle
spectra, see~\cite{stecker2007}. This implies, the exact spectral index
depends on the shock properties. As we are not dealing with observed
GRBs here, but with potentially to-be-observed GRBs, we cannot
determine the spectral behavior by the observation of the synchrotron
spectrum as it was done in~\cite{guetta2004,bshr2006}. Thus, we use
\mbox{$\alpha_{\nu}=1$} and \mbox{$\beta_{\nu}=2$} as a first order approximation, just as
it is done in several approaches of unknown single-source spectra, see e.g.~\citep{wb1997,wb1999}. 
\item {\it The normalization $A_{\nu}$}\\
The normalization, which corresponds to the intensity of the burst, is
given by
\begin{equation}
A_{\nu}=\frac{1}{8}\frac{1}{f_e}\frac{E_{\gamma}^{iso}}{4\pi\cdot
  d_{L}^{2}}\cdot \frac{f_{\pi}}{\ln(10)}\,.
\label{anu:equ}
\end{equation}
Here, we use a fixed isotropic energy release of
\mbox{$E_{\gamma}^{iso}=L_{\gamma}\cdot t_{90}=10^{52}$ erg}, where we assume a burst
duration of \mbox{$t_{90}\approx 10$ s}. Further, we assume that the energy contained
in GRB electrons is $1/10$th of the energy contained in GRB
protons, \mbox{$f_{e}=0.1$}. The factor  \mbox{$f_{\pi}\approx 0.2$} describes the fraction of
energy transferred to the charged pion. The factor $1/8$ is applied since $1/2$ of all
proton-photon interactions go into neutrino production, and $1/4$ goes
into a single neutrino flavor. Here, we fix all parameters except for
the distance $d_L$ of the GRB from Earth, which is determined by the
distance of the starburst from Earth.
\end{itemize}
To estimate the neutrino flux from a standard GRB from one of the starburst
galaxies in our sample, we simply calculate a normalization, dependent on the
distance of the starburst. Every other parameter is kept constant. Hence, the
results can only serve as a rough estimate of what would happen on
average. Both the spectral indices, the break energies and the normalization
fluctuate with each burst as described theoretically
in~\cite{guetta2004,bshr2006} and worked out experimentally in AMANDA for the
case of GRB030329~\citep{mike_icrc2005} and in IceCube for
GRB080319B~\citep{alexander_grb080319B_conf}. One of the most important parameters is the total GRB energy. Here,
we assume that it is \mbox{$E_{\gamma}^{iso}=10^{52}$ erg}. The isotropic equivalent energy for the 'naked eye' GRB080319B is two orders of
magnitude higher,
i.e.\ \mbox{$1.3\cdot 10^{54}$ erg} and several weak bursts will have smaller isotropic
equivalent energies. While these effects cannot be taken into account due to
lack of knowledge, they should be kept in mind. In case of a positive detection
in a high-energy neutrino detector, the parameters can of course be determined
explicitly. In the case of a negative result, it would be possible to set a
limit in the following sense: no GRB with isotropic equivalent luminosities
occurred during the time of observation in the given starburst(s).
\begin{figure}
\centering{
\epsfig{file=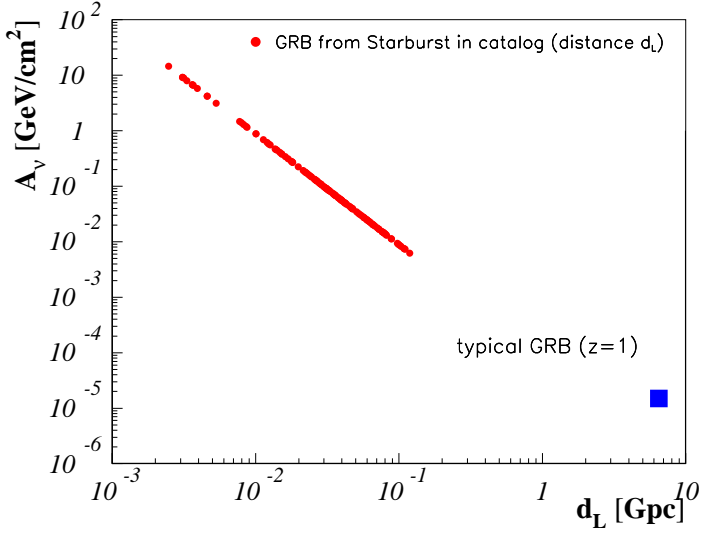,width=10cm}
\caption{Neutrino intensity $A_{\nu}$ for a single GRB from a starburst at a
  luminosity distance $d_L$. A generic GRB isotropic energy of
  \mbox{$E_{\gamma}^{iso}=10^{52}$ erg}
  is assumed. Therefore, the neutrino intensity only depends on the
  luminosity distance as  \mbox{$A_{\nu}\propto d_{l}^{-2}$}.\label{cat_anu_z:fig}} 
}
\end{figure}

\begin{figure}
\centering{
\epsfig{file=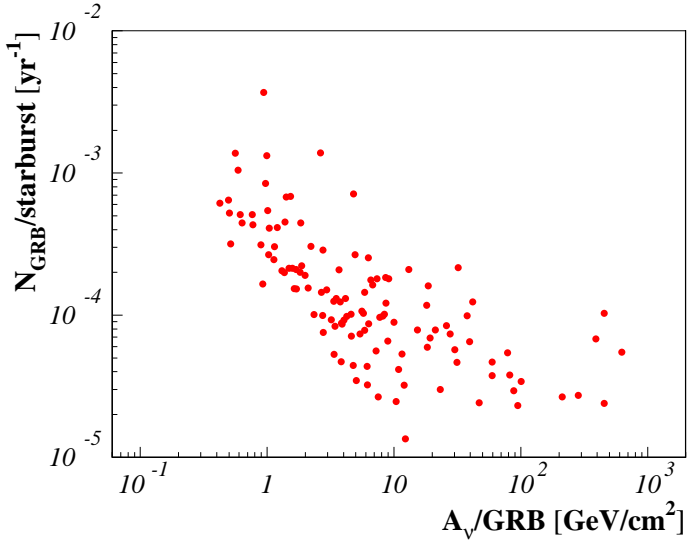,width=10cm}
\caption{Number of GRBs in a starburst versus neutrino flux
  intensity for each starburst. The sharp cutoff at low neutrino
  intensities comes from the distance cut performed at $z=0.03$. This
  leads to a cutoff in the neutrino intensity, since the latter is
  proportional to the luminosity distance squared. \label{nrgrbs_anu:fig}} 
}
\end{figure}
\paragraph{Expected events in IceCube}
The probability of a GRB to be observed from one of the starburst galaxies in
the sample from the northern hemisphere is shown in Fig.~\ref{grbs_year:fig} (blue squares).
Analyzing all $96$ sources in the northern
hemisphere, a rate of 
\be
\dot{n}_{\rm GRB}^{\rm total/north}(N_{\rm starbursts}=96)=0.02\,{\rm yr}^{-1}
\ee 
is expected. This number can be enhanced significantly when we
consider all starbursts originally selected (309 sources). We
only consider the brightest ones for completeness reasons. Thus, we
expect that at least one GRB from a starburst in the supergalactic
plane should happen in the lifetime of IceCube, if not more. 
 Prospects for Km3NeT are not as optimal as for IceCube,
since Km3NeT will be located in the northern hemisphere, looking at
the southern sky, and the instrument will therefore
only see a small fraction of the class of starburst galaxies, which
dominantly shows sources in the northern hemisphere.

\begin{figure}
\centering{
\epsfig{file=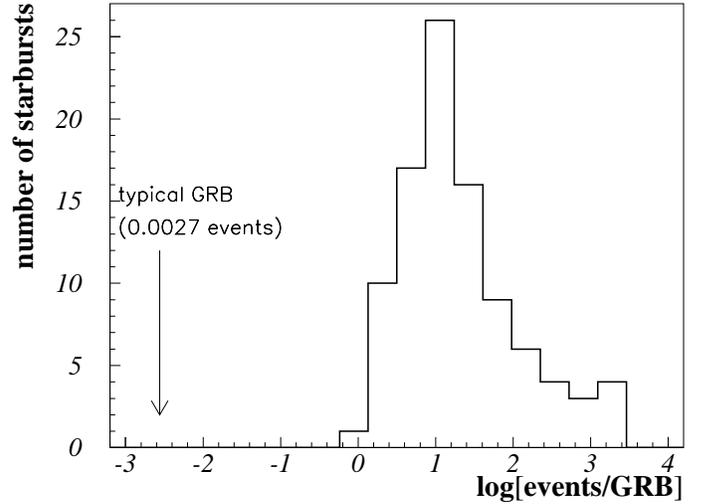,width=10cm}
\caption{Histogram of number of events in IceCube from the 96 starburst
  galaxies in the northern hemisphere. The IceCube effective area we use
  \citep{montaruli_taup2007} does not take into account sensitivity
  changes with declination. All bursts are stronger than 1 event in
  IceCube, since they come from starbursts closer than $z=0.03$. For more distant bursts, the number of events will be
  less, since the neutrino intensity decreases with the distance squared.\label{icecube_eventrate:fig}} 
}
\end{figure}

IceCube's effective area $A_{eff}$ is presented by~\cite{montaruli_taup2007} and can be used to
determine the total number of events per GRB in IceCube, $N_{events}$, by folding it with the
GRB spectrum $dN_{\nu}/d\en$,
\begin{equation}
N_{events}=\int_{E_{th}}^{\infty} A_{eff}(\en)\cdot \frac{dN_{\nu}}{d\en}(\en)\, d\en\,.
\end{equation}
Here, we neglect the weak dependence of the effective area on the declination
of the burst. For the threshold energy, we use \mbox{$E_{th}=100$ GeV}
\citep{icecube_performance2004}. This is the
general detection threshold of IceCube. As events can be selected by direction
and a small time window, the atmospheric background can be reduced to almost
zero. Therefore, events in the entire energy range are available in such an
analysis. If we now assume a standard burst with an isotropic energy release
of \mbox{$E_{\gamma}^{iso}=10^{52}$ erg}, we can estimate how strong a GRB would be
on average from the nearby starbursts in our
catalog. Figure~\ref{icecube_eventrate:fig} shows the histogram of the number
of events per starburst for an average GRB. Depending on the distance of the
starburst, the event rate in IceCube ranges from more than 1 event up to more
than 1000 events per burst in a small time window of around $10-100$
seconds. These numbers lie between 2 and 6 orders of magnitude above those
GRBs typically observed by satellite experiments like BATSE, Swift and
Fermi. The reason why such a strong burst has not been observed in gamma-rays
is simply the relatively low rate of occurrence\footnote{The exact number
cannot be given here, since we only consider those sources with
relatively high fluxes in the radio and at FIR wavelengths. The total
rate of 0.02 GRBs/year could be enhanced significantly with a lower
flux threshold.}. If such an event happens, neutrino detectors, and also
cosmic ray detectors have a higher chance of detection due to their extremely
large FoV ($\sim 2\pi$).

\subsubsection{High-energy photons and nearby GRBs}
Following the channel of $\pi^{0}$ production rather than the
neutrino-producing charged pions, we see that a high-energy photon
signal is always expected along with high-energy neutrino
emission. Such a correlation could be observed for sources optically
thin to TeV photons. In such a case, GRBs in nearby starburst galaxies
would produce an equally strong signal in neutrinos as in high-energy
photons. This means a good discovery potential for large FoV detectors
at GeV-TeV energies. Fermi is one of the best candidates to observe
this high-energy radiation at GeV energies, Milagro and its successor
HAWC will be able to see such events at TeV energies. Air Cherenkov
Telescopes like H.E.S.S., MAGIC and VERITAS have a FoV that is too
small and are therefore not well-suited for a detection of those rare
GRBs. If, however, such a GRB is detected by Swift, especially MAGIC
will be able to follow up the GRB quickly, as the telescope is
designed for quick GRB follow-ups, see e.g.~\cite{galante2008}.
\subsubsection{Summary}
To sum up, if a GRB will happen in one of the 127 starburst galaxies,
the detection probability is extremely high in wide-FoV detectors for
 high-energy neutrinos and photons as well as in cosmic rays. Since
 most sources are located in the northern hemisphere, detectors like
 IceCube, HAWC and
 Auger are optimal for for such a study. This
 would enable a detailed study of the hadronic emission processes of
 GRBs, since the main three messenger particles are covered. Explicit search methods for an enhancement of
cosmic rays, high-energy photons and neutrinos from those starbursts can clarify weather or not such a burst
happens or not. In the case of the closest sources, the GRB should even be
visible in a typical skymap of neutrino, photon or cosmic ray telescopes.

\section{Implications and possible experimental tests\label{implications}}
In this paper, we present a model to explain the correlation between
far-infrared and radio emission in starburst galaxies and we discuss
the particle spectra from cosmic ray interactions and possibilities of
their observation at Earth. In this context, we also present a catalog
of 127 nearby ($z<0.03$) starburst galaxies with both FIR ($\ssixty>4$~Jy) and
radio ($\sone>20$~mJy) measurements.
Those starbursts lie in the supergalactic plane ($z<0.03$) and serve
as a test for basic properties of starbursts. It can further be used to
calculate the cosmic ray emission from starbursts in the local Universe. In
this section, we summarize implications:
\begin{enumerate}
\item {\it FIR-Radio correlation}\\
Within this model, we show that the radio emission is basically
independent of the magnetic field strength. This leads to a
calorimetric correlation between radio and FIR emission.

Starburst galaxies as well as quiescent galaxies equally obey the correlation
between the radio and the far-infrared emission.  Therefore the theory
implies by necessity that the magnetic field in a starburst region rises with
the overall energy density. Since galactic dynamos are not understood, this just
raises the ante in terms of the requirements for any theory that tries to
account for the magnetic fields and their strengths in galaxies, see \citep[e.g.]{beck_hoernes1996}.

Interacting galaxies, subject to substantial compression, show magnetic fields higher than corresponds
to the energy density derived from star formation. Therefore, it may be
expected that they have more radio emission than indicated by the general radio/FIR
correlation. At least one example shows this kind of behavior \citep{hummel_beck1995}.

There is likely also a lowest level of star formation activity, where the
assumption that the hot medium is fully connected, fails.  There one may expect
also substantial departures from the correlation.

\item {\it High-energy photons and neutrinos from SNRs in starbursts}\\
The diffuse high-energy neutrino and photon flux from SNRs in
starbursts has been discussed
previously by \cite{lw2006,thompson2007}. In our calculation, we
assume that a fraction of the total energy of a SNR is transferred to
cosmic ray acceleration. It is usually assumed that supernova remnants in the Galaxy accelerate
particles up to the knee, i.e. $E_{\max} \sim 10^{15}$~eV. Under the
assumption that this is the maximum energy for SNRs in starburst
galaxies, we show that a detection of the diffuse neutrino flux from
starbursts with IceCube or Km3NeT is unlikely. The main reason is the
low energy cutoff, and not the strength of the signal. On the
other hand, if it were
possible to accelerate protons to even higher energies in SNRs in
starbursts, as is suggested by \cite{lw2006}, IceCube will be
sensitive to the neutrino flux. High-energy photons make up around
$10\%$ of the total high-energy photon background observed by
EGRET. The high-energy photon detector Fermi was launched on June 11,
2008 and will soon be able to give more details on the background and
which sources and source classes contribute. Cosmic rays themselves
cannot be observed from starburst galaxies, since the background of
Galactic cosmic rays is too high. 
\item {\it High-energy neutrinos and cosmic rays from GRBs in nearby
  starbursts}\\
Due to an enhanced rate of star formation and supernova explosions in
starbursts, these Galaxies provide a higher probability for Gamma Ray
Bursts compared to regular galaxies. So far, Gamma Ray Bursts could
not be identified in high-energy neutrinos. Despite their tremendous
release of neutrinos, they are usually too far away to yield a
significant signal individually ($z\sim 1-2$). The stacking of known GRB
locations in neutrino data was done with the AMANDA experiment, see
\cite{casc_paper2007,kyler2008}, with limits above the typical, expected diffuse
flux. IceCube will be able to explore the diffuse neutrino flux using
the stacking method. Two single GRBs were also analysed, the first one
being GRB030329, using AMANDA data \citep{mike_icrc2005}, the second one
being GRB080319B, using IceCube in a 9-string configuration \citep{alexander_grb080319B_conf}.  
In this paper, we show using our catalog of starbursts in the
supergalactic plane, that GRBs in closeby starbursts ($z<0.03$) typically
produce event rates in IceCube above 1 event per bursts. For the most
distant sources at $z=0.03$, about 1 event is expected, for the
closest sources, more than 1000 events are predicted within a short
time window of a few tens of seconds. In our catalog of 127
starbursts, about 1 GRB is expected each 30th year, most of those
happening in the northern hemisphere. The total rate of GRBs in the
supergalactic plane is, however, expected to be higher, since we only
include the brightest sources in this catalog ($\ssixty>4$~Jy and $\sone>20$~mJy).
The stacking of starburst
galaxies for a neutrino analysis would help identifying the sources at
$z\sim 0.03$, while the closest sources will immediately be visible in
a neutrino skymap. Those GRBs should be accompanied by high-energy photon
emission as well. Fermi will be able to identify these at GeV
energies \citep{glast_gamma_santafe2007}. At even higher energies, Milagro \citep{milagro_gamma_santafe2007} and the next generation
experiment HAWC \citep{hawc,hawc_madison06} will be able to investigate TeV emission. For such
nearby and strong cosmic ray emitters, it might even be possible to
identify them in charged cosmic rays, given that diffusion in time and
space is not too large.
\end{enumerate}

\begin{acknowledgements}
Chapter 4 was distributed as an MPIfR preprint in an earlier
version 1995, following a lecture at a conference in Heidelberg.
We would like to thank W.~I.~Axford, R.~Beck, A.~Bell, E.~Berkhuijsen,
D.~Breitschwerdt, R.~Chini, K.~Chy\.zy, L.~Drury, C.~Galea,
F.~Halzen, M.~Harwit, R.~Jokipii, A.~Kandus, H.~Kang, A.~Kappes,
U.~Klein, T.~de Jong, P.~P.~Kronberg, H.~S.~Lee, J.~Mathis,
W.~M.~Matthaeus, H.~Meyer, P.~G.~Mezger, K.~Otmianowska-Mazur, V.~Ptuskin, M. and G.~Rieke, W.~Rhode, 
D.~Ryu, T.~Schmutzler$^{\dag}$, E.-S.~Seo, L.~F.~Smith, T.~Stanev, F.~Tabatabaei, M.~Tjus,
U.~Torkelsson, M.~Urbanik, H.-J.~V{\"o}lk, E.~Waxman, J.~Wefel,
B.~Wiebel-Sooth, R.~Wielebinski and C.~Wiebusch for inspiring
discussions. 
JKB and
JD would like to thank the IceCube collaboration for useful
comments. 
Support for PLB is coming from the
AUGER membership and theory grant 05 CU 5PD 1/2 via DESY/BMBF, as well
as from VIHKOS. Support for JKB is coming from the
DFG grant BE-3714/3-1. Further, JD and JKB are supported by the IceCube grants
BMBF (05 CI5PE1/0) and (05 A08PE1). Support for TMK is coming
  from the DFG grant Kn~765/1-2.
\end{acknowledgements}
\begin{appendix}
\section{Catalog tables\label{appendix_a}}
We present the catalog of nearby starburst galaxies
discussed throughout the paper. The 127 sources presented here were
selected from a larger sample of 309 sources, all previously
identified as starburst galaxies. We require the FIR flux at $60\,\mu$ to be larger than
$\ssixty>4$~Jy and the radio flux at $1.4$~GHz to be larger than
$\sone>20$~mJy. In addition, we only include sources at
$z<0.03$. Table \ref{genfir:tab} summarizes the basic properties of the catalog:
name, right ascension (RA [deg]), declination (DEC [deg]), redshift
(z) and distance ($\text{D}_\text{L}$ [Gpc]), together with the FIR flux measurements, $S_{\lambda}$, $\lambda$ giving the
wavelength. Table \ref{radio:tab} presents radio flux measurements,
$S_{\nu}$, with $\nu$ as the frequency. Table \ref{xray:tab}
summarizes X-ray flux measurements. References are always given in the
last column.  
\longtabL{1}{
\begin{landscape}
\begin{longtable}{|l|c|c|c|c|c|c|c|c|c|c|}
\caption{\label{genfir:tab}Coordinates, distances and IRAS
measurements of the sample. All fluxes in [Jy]. Coordinates epoch
J2000.0 and distances are obtained from NED.  In NED, distances are
corrected to the cosmological microwave background using cosmological
parameters according to $\Lambda$CDM cosmology,
$h_0=0.73,\,\Omega_{m}=0.27,\,\Omega_{\Lambda}=0.73$, using \cite{2006PASP..118.1711W}. \\
IRAS References:\\
1:~\cite{2003AJ....126.1607S}, 2:~\cite{1990irasf_c_0000m}, 3:~\cite{2007A&A...462..507L}, 4:~\cite{2004AJ....127.3235S}, 5:~\cite{1994PrivC.U..J....K}, 6:~\cite{1988iras....1.....B}, 7:~\cite{1989AJ.....98..766S}}\\
\hline
\textbf{Name}& \textbf{RA [deg]} & \textbf{DEC [deg]} & \textbf{z} & \textbf{$\text{D}_\text{L} [Gpc]$} & $\mathbf{S_{12\,\mu m}}$ & $\mathbf{S_{25\,\mu m}}$ & $\mathbf{S_{60\,\mu m}}$ & $\mathbf{S_{100\,\mu m}}$ & \textbf{References}\\
\hline\hline
\endfirsthead
\caption{continued.}\\
\hline
\textbf{Name}& \textbf{RA [deg]} & \textbf{DEC [deg]} & \textbf{z} & \textbf{$\text{D}_\text{L} [Gpc]$} & $\mathbf{S_{12\,\mu m}}$ & $\mathbf{S_{25\,\mu m}}$ & $\mathbf{S_{60\,\mu m}}$ & $\mathbf{S_{100\,\mu m}}$ & \textbf{References}\\
\hline\hline
\endhead
\hline
\endfoot
MRK545 & $2.47254$ & $25.9238$ & $0.01523$ & $0.05962$ & $0.523$ & $1.082$ & $9.196$ & $15.34$ & $2$\\
NGC34 & $2.77729$ & $-12.1073$ & $0.019617$ & $0.0771$ & $0.35$ & $2.39$ & $17.05$ & $16.86$ & $1$\\
MCG-02-01-051 & $4.71202$ & $-10.3768$ & $0.027103$ & $0.109$ & $0.24$ & $1.19$ & $7.35$ & $10.22$ & $4$\\
NGC174 & $9.24558$ & $-29.4778$ & $0.011905$ & $0.0451$ & $0.41$ & $1.27$ & $11.36$ & $19.77$ & $1$\\
NGC232 & $10.6909$ & $-23.5614$ & $0.022172$ & $0.0886$ & $0.36$ & $1.28$ & $10.05$ & $17.14$ & $1$\\
NGC253 & $11.888$ & $-25.2882$ & $0.0008$ & $0.0031$ & $41.04$ & $154.67$ & $967.81$ & $1288.15$ & $1$\\
IC1623 & $16.9466$ & $-17.507$ & $0.02007$ & $0.07857$ & $1.03$ & $3.65$ & $22.93$ & $31.55$ & $1$\\
NGC520 & $21.1461$ & $3.79242$ & $0.00761$ & $0.03022$ & $0.9$ & $3.22$ & $31.52$ & $47.37$ & $2$\\
NGC632 & $24.323$ & $5.87764$ & $0.010567$ & $0.0396$ & $0.37$ & $0.88$ & $4.89$ & $7.32$ & $5$\\
NGC660 & $25.7598$ & $13.6457$ & $0.00283$ & $0.01233$ & $3.05$ & $7.3$ & $65.52$ & $114.74$ & $1$\\
NGC828 & $32.5399$ & $39.1904$ & $0.01793$ & $0.07073$ & $0.72$ & $1.07$ & $11.46$ & $25.33$ & $1$\\
NGC891 & $35.6392$ & $42.3491$ & $0.00176$ & $0.00857$ & $5.27$ & $7$ & $66.46$ & $172.23$ & $1$\\
NGC958 & $37.6785$ & $-2.939$ & $0.01914$ & $0.0765$ & $0.62$ & $0.94$ & $5.85$ & $15.08$ & $1$\\
NGC1055 & $40.4385$ & $0.443167$ & $0.00332$ & $0.01131$ & $2.24$ & $2.84$ & $23.37$ & $65.26$ & $1$\\
Maffei2 & $40.4795$ & $59.6041$ & $-5.7e-05$ & $0.00332$ & $3.624$ & $9.238$ & $135$ & $225$ & $6$\\
NGC1068(M77) & $40.6696$ & $-0.0132806$ & $0.00379$ & $0.0137$ & $39.84$ & $87.57$ & $196.37$ & $257.37$ & $1$\\
UGC2238 & $41.5729$ & $13.0957$ & $0.021883$ & $0.0883$ & $0.36$ & $0.65$ & $8.17$ & $15.67$ & $1$\\
NGC1097 & $41.5794$ & $-30.2749$ & $0.00424$ & $0.0152$ & $2.96$ & $7.3$ & $53.35$ & $104.79$ & $1$\\
NGC1134 & $43.4222$ & $13.0141$ & $0.012142$ & $0.0474$ & $0.55$ & $0.92$ & $9.09$ & $19.43$ & $1$\\
NGC1365 & $53.4015$ & $-36.1404$ & $0.00546$ & $0.01793$ & $5.12$ & $14.28$ & $64.31$ & $165.67$ & $1$\\
IC342 & $56.7021$ & $68.0961$ & $0.0001$ & $0.0046$ & $14.92$ & $34.48$ & $180.8$ & $391.66$ & $1$\\
UGC02982 & $63.0935$ & $5.54739$ & $0.017696$ & $0.0724$ & $0.57$ & $0.83$ & $8.39$ & $16.82$ & $1$\\
NGC1530 & $65.8629$ & $75.2956$ & $0.00821$ & $0.03622$ & $0.72$ & $1.23$ & $9.88$ & $25.88$ & $1$\\
NGC1569 & $67.7044$ & $64.8479$ & $-0.00035$ & $0.0046$ & $1.24$ & $9.03$ & $54.36$ & $55.29$ & $1$\\
MRK617 & $68.4994$ & $-8.57888$ & $0.01594$ & $0.06261$ & $0.441$ & $7.286$ & $32.31$ & $32.69$ & $1$\\
NGC1672 & $71.4271$ & $-59.2473$ & $0.00444$ & $0.01682$ & $2.47$ & $5.25$ & $41.21$ & $77.92$ & $1$\\
MRK1088 & $73.6598$ & $3.26797$ & $0.01528$ & $0.06051$ & $0.2659$ & $0.835$ & $6.605$ & $10.77$ & $1$\\
NGC1808 & $76.9264$ & $-37.5131$ & $0.00332$ & $0.01261$ & $5.4$ & $17$ & $105.55$ & $141.76$ & $1$\\
NGC1797 & $76.937$ & $-8.01908$ & $0.014814$ & $0.0616$ & $0.33$ & $1.35$ & $9.56$ & $12.76$ & $1$\\
MRK1194 & $77.9423$ & $5.20061$ & $0.01491$ & $0.05948$ & $0.283$ & $0.7071$ & $6.688$ & $11.5$ & $2$\\
NGC2146 & $94.6571$ & $78.357$ & $0.00298$ & $0.012$ & $6.83$ & $18.81$ & $146.69$ & $194.05$ & $1$\\
NGC2276 & $111.81$ & $85.7546$ & $0.00804$ & $0.0328$ & $1.07$ & $1.63$ & $14.29$ & $28.97$ & $1$\\
NGC2403 & $114.214$ & $65.6026$ & $0.00044$ & $0.00247$ & $2.82$ & $3.57$ & $41.47$ & $99.13$ & $1$\\
NGC2415 & $114.236$ & $35.242$ & $0.01262$ & $0.05341$ & $0.61$ & $1.19$ & $8.75$ & $13.58$ & $1$\\
NGC2782 & $138.521$ & $40.1137$ & $0.00848$ & $0.03951$ & $0.64$ & $1.51$ & $9.17$ & $13.76$ & $1$\\
NGC2785 & $138.814$ & $40.9175$ & $0.008746$ & $0.0392$ & $0.49$ & $1.09$ & $8.4$ & $15.79$ & $1$\\
NGC2798 & $139.346$ & $41.9997$ & $0.00576$ & $0.02784$ & $0.76$ & $3.21$ & $20.6$ & $29.69$ & $2$\\
NGC2903 & $143.042$ & $21.5008$ & $0.00186$ & $0.00826$ & $5.29$ & $8.64$ & $60.54$ & $130.43$ & $1$\\
MRK708 & $145.548$ & $4.67314$ & $0.00682$ & $0.03116$ & $0.46$ & $0.8$ & $5.36$ & $8.24$ & $1$\\
NGC3034(M82) & $148.968$ & $69.6797$ & $0.00068$ & $0.00363$ & $79.43$ & $332.63$ & $1480.42$ & $1373.69$ & $1$\\
NGC3079 & $150.491$ & $55.6797$ & $0.00375$ & $0.01819$ & $2.54$ & $3.61$ & $50.67$ & $104.69$ & $1$\\
NGC3147 & $154.224$ & $73.4007$ & $0.00941$ & $0.04141$ & $1.95$ & $1.03$ & $8.17$ & $29.61$ & $1$\\
NGC3256 & $156.964$ & $-43.9038$ & $0.00935$ & $0.03535$ & $3.57$ & $15.69$ & $102.63$ & $114.31$ & $1$\\
MRK33 & $158.133$ & $54.401$ & $0.00477$ & $0.0221$ & $0.21$ & $1.05$ & $4.77$ & $5.99$ & $5$\\
NGC3310 & $159.691$ & $53.5034$ & $0.00331$ & $0.01981$ & $1.54$ & $5.32$ & $34.56$ & $44.19$ & $1$\\
NGC3367 & $161.646$ & $13.7509$ & $0.010142$ & $0.0468$ & $0.51$ & $1.98$ & $6.44$ & $13.48$ & $1$\\
NGC3448 & $163.663$ & $54.3052$ & $0.0045$ & $0.02406$ & $0.22$ & $0.64$ & $6.64$ & $11.17$ & $1$\\
NGC3504 & $165.797$ & $27.9725$ & $0.00512$ & $0.02707$ & $1.11$ & $4.03$ & $21.43$ & $34.05$ & $1$\\
NGC3556(M108) & $167.879$ & $55.6741$ & $0.00233$ & $0.01385$ & $2.29$ & $4.19$ & $32.55$ & $76.9$ & $1$\\
NGC3627(M66) & $170.063$ & $12.9915$ & $0.00243$ & $0.01004$ & $4.82$ & $8.55$ & $66.31$ & $136.56$ & $1$\\
NGC3628 & $170.071$ & $13.5895$ & $0.00281$ & $0.01004$ & $3.13$ & $4.85$ & $54.8$ & $105.76$ & $1$\\
NGC3683 & $171.883$ & $56.8771$ & $0.005724$ & $0.0259$ & $1.16$ & $1.48$ & $13.87$ & $29.3$ & $1$\\
NGC3690 & $172.134$ & $58.5622$ & $0.01041$ & $0.04774$ & $3.97$ & $24.51$ & $113.05$ & $111.42$ & $2$\\
MRK188 & $176.893$ & $55.9672$ & $0.00803$ & $0.0355$ & $0.3621$ & $0.4515$ & $4.576$ & $11.52$ & $2$\\
NGC3893 & $177.159$ & $48.7108$ & $0.00323$ & $0.0161$ & $1.45$ & $1.65$ & $15.57$ & $36.8$ & $1$\\
NGC3994 & $179.404$ & $32.2776$ & $0.010294$ & $0.0466$ & $0.32$ & $0.46$ & $4.98$ & $10.31$ & $4$\\
NGC4030 & $180.099$ & $-1.1$ & $0.00487$ & $0.0245$ & $1.35$ & $2.3$ & $18.49$ & $50.92$ & $1$\\
NGC4041 & $180.551$ & $62.1373$ & $0.00412$ & $0.02278$ & $1.13$ & $1.56$ & $14.15$ & $31.74$ & $2$\\
NGC4102 & $181.596$ & $52.7109$ & $0.002823$ & $0.0141$ & $1.77$ & $6.83$ & $46.85$ & $70.29$ & $1$\\
MRK1466 & $182.046$ & $2.87828$ & $0.00443$ & $0.01529$ & $0.325$ & $1.236$ & $6.265$ & $10.52$ & $2$\\
MRK759 & $182.656$ & $16.0329$ & $0.00723$ & $0.0345$ & $0.2995$ & $0.537$ & $4.116$ & $8.727$ & $2$\\
NGC4194 & $183.539$ & $54.5268$ & $0.00834$ & $0.04033$ & $0.99$ & $4.51$ & $23.2$ & $25.16$ & $1$\\
NGC4214 & $183.913$ & $36.3269$ & $0.00097$ & $0.00367$ & $0.58$ & $2.46$ & $17.57$ & $29.08$ & $2$\\
NGC4273 & $184.984$ & $5.34331$ & $0.007932$ & $0.0376$ & $0.77$ & $1.65$ & $9.38$ & $21.76$ & $1$\\
NGC4303(M62) & $185.479$ & $4.47365$ & $0.005224$ & $0.0264$ & $3.28$ & $4.9$ & $37.27$ & $78.74$ & $1$\\
NGC4414 & $186.613$ & $31.2235$ & $0.00239$ & $0.01768$ & $2.78$ & $3.61$ & $29.55$ & $70.69$ & $1$\\
NGC4418 & $186.728$ & $-0.877556$ & $0.007268$ & $0.0349$ & $0.99$ & $9.67$ & $43.89$ & $31.97$ & $1$\\
NGC4527 & $188.535$ & $2.65381$ & $0.005791$ & $0.0286$ & $2.65$ & $3.55$ & $31.4$ & $65.68$ & $1$\\
NGC4536 & $188.613$ & $2.18789$ & $0.006031$ & $0.0297$ & $1.55$ & $4.04$ & $30.26$ & $44.51$ & $1$\\
NGC4631 & $190.533$ & $32.5415$ & $0.00202$ & $0.00773$ & $5.16$ & $8.97$ & $85.4$ & $160.08$ & $1$\\
NGC4666 & $191.286$ & $-0.461885$ & $0.005101$ & $0.0257$ & $3.34$ & $3.89$ & $37.11$ & $85.95$ & $1$\\
NGC4793 & $193.67$ & $28.9383$ & $0.008286$ & $0.038$ & $1.08$ & $1.57$ & $12.42$ & $28.11$ & $1$\\
NGC4826(M64) & $194.182$ & $21.6811$ & $0.00136$ & $0.0309$ & $2.36$ & $2.86$ & $36.7$ & $81.65$ & $1$\\
NGC4945 & $196.364$ & $-49.4682$ & $0.00187$ & $0.00392$ & $27.47$ & $42.34$ & $625.46$ & $1329.7$ & $1$\\
NGC5005 & $197.734$ & $37.0592$ & $0.00316$ & $0.01809$ & $1.65$ & $2.26$ & $22.18$ & $63.4$ & $1$\\
NGC5020 & $198.166$ & $12.5998$ & $0.011214$ & $0.0507$ & $0.36$ & $0.72$ & $5.58$ & $11.7$ & $1$\\
NGC5055(M63) & $198.956$ & $42.0293$ & $0.00168$ & $0.00796$ & $5.35$ & $6.36$ & $40$ & $139.82$ & $1$\\
ARP193 & $200.147$ & $34.1395$ & $0.02335$ & $0.101$ & $0.25$ & $1.42$ & $17.04$ & $24.38$ & $1$\\
NGC5104 & $200.346$ & $0.342417$ & $0.018606$ & $0.082$ & $0.39$ & $0.74$ & $6.78$ & $13.37$ & $1$\\
NGC5135 & $201.434$ & $-29.8337$ & $0.01372$ & $0.05215$ & $0.63$ & $2.38$ & $16.86$ & $30.97$ & $1$\\
NGC5194(M51) & $202.47$ & $47.1952$ & $0.00154$ & $0.00873$ & $7.21$ & $9.56$ & $97.42$ & $221.21$ & $1$\\
NGC5218 & $203.043$ & $62.7678$ & $0.009783$ & $0.0419$ & $0.37$ & $0.94$ & $7.01$ & $13.54$ & $1$\\
NGC5236(M83) & $204.254$ & $-29.8657$ & $0.00172$ & $0.00363$ & $21.46$ & $43.57$ & $265.84$ & $524.09$ & $1$\\
NGC5256 & $204.573$ & $48.2769$ & $0.027863$ & $0.119$ & $0.32$ & $1.07$ & $7.25$ & $10.11$ & $1$\\
NGC5257 & $204.968$ & $0.839583$ & $0.022676$ & $0.099$ & $0.52$ & $1.18$ & $8.1$ & $13.63$ & $4$\\
NGC5253 & $204.983$ & $-31.6401$ & $0.00136$ & $0.00315$ & $2.612$ & $12.07$ & $29.84$ & $30.08$ & $2$\\
UGC8739 & $207.308$ & $35.2574$ & $0.016785$ & $0.0728$ & $0.35$ & $0.42$ & $5.79$ & $15.89$ & $1$\\
MRK1365 & $208.63$ & $15.0441$ & $0.01846$ & $0.0806$ & $0.1562$ & $0.6445$ & $4.203$ & $6.113$ & $2$\\
NGC5430 & $210.191$ & $59.3283$ & $0.009877$ & $0.0423$ & $0.5$ & $1.94$ & $10.1$ & $20.34$ & $1$\\
NGC5427 & $210.859$ & $-6.03081$ & $0.008733$ & $0.0399$ & $1.29$ & $1.48$ & $10.24$ & $25.29$ & $1$\\
NGC5678 & $218.023$ & $57.9214$ & $0.00641$ & $0.03202$ & $0.94$ & $1.2$ & $9.67$ & $25.66$ & $1$\\
NGC5676 & $218.195$ & $49.4579$ & $0.007052$ & $0.0308$ & $1.13$ & $1.7$ & $12.04$ & $29.91$ & $1$\\
NGC5713 & $220.048$ & $-0.289222$ & $0.00658$ & $0.02674$ & $1.47$ & $2.84$ & $22.1$ & $37.28$ & $1$\\
NGC5775 & $223.49$ & $3.54446$ & $0.00561$ & $0.02634$ & $1.83$ & $2.47$ & $23.59$ & $55.64$ & $1$\\
NGC5900 & $228.772$ & $42.2094$ & $0.008376$ & $0.0361$ & $0.4$ & $0.7$ & $7.51$ & $16.95$ & $1$\\
NGC5936 & $232.504$ & $12.9893$ & $0.013356$ & $0.0575$ & $0.48$ & $1.47$ & $8.73$ & $17.66$ & $1$\\
ARP220 & $233.738$ & $23.5032$ & $0.01813$ & $0.0799$ & $0.61$ & $8$ & $104.09$ & $115.29$ & $1$\\
NGC5962 & $234.132$ & $16.6079$ & $0.006528$ & $0.0288$ & $0.73$ & $1.04$ & $8.93$ & $21.82$ & $1$\\
NGC5990 & $236.568$ & $2.41547$ & $0.012806$ & $0.055$ & $0.6$ & $1.6$ & $9.59$ & $17.14$ & $1$\\
NGC6181 & $248.087$ & $19.8266$ & $0.007922$ & $0.0334$ & $0.63$ & $1.41$ & $8.94$ & $20.83$ & $1$\\
NGC6217 & $248.163$ & $78.1982$ & $0.00454$ & $0.02349$ & $0.74$ & $2.03$ & $11.35$ & $20.62$ & $1$\\
NGC6240 & $253.245$ & $2.40094$ & $0.02448$ & $0.10336$ & $0.59$ & $3.55$ & $22.94$ & $26.49$ & $1$\\
NGC6286 & $254.631$ & $58.9363$ & $0.018349$ & $0.0761$ & $0.47$ & $0.62$ & $9.24$ & $23.11$ & $1$\\
IRAS18293-3413 & $278.171$ & $-34.191$ & $0.01818$ & $0.07776$ & $1.14$ & $3.98$ & $35.71$ & $53.38$ & $1$\\
NGC6701 & $280.802$ & $60.6533$ & $0.01323$ & $0.05664$ & $0.55$ & $1.32$ & $10.05$ & $20.05$ & $1$\\
NGC6764 & $287.068$ & $50.9332$ & $0.008059$ & $0.03131$ & $0.54$ & $1.33$ & $6.62$ & $12.44$ & $1$\\
NGC6946 & $308.718$ & $60.1539$ & $0.00016$ & $0.00532$ & $12.11$ & $20.7$ & $129.78$ & $290.69$ & $1$\\
NGC7130 & $327.081$ & $-34.9513$ & $0.01615$ & $0.06599$ & $0.58$ & $2.16$ & $16.71$ & $25.89$ & $1$\\
IC5179 & $334.038$ & $-36.8437$ & $0.01141$ & $0.0467$ & $1.18$ & $2.4$ & $19.39$ & $37.29$ & $1$\\
NGC7331 & $339.267$ & $34.4156$ & $0.00272$ & $0.01471$ & $3.94$ & $5.92$ & $45$ & $110.16$ & $1$\\
NGC7469 & $345.815$ & $8.874$ & $0.01632$ & $0.06523$ & $1.59$ & $5.96$ & $27.33$ & $35.16$ & $1$\\
NGC7479 & $346.236$ & $12.3229$ & $0.00794$ & $0.03236$ & $1.37$ & $3.86$ & $14.93$ & $26.73$ & $1$\\
NGC7496 & $347.447$ & $-43.4279$ & $0.0055$ & $0.02234$ & $0.58$ & $1.93$ & $10.14$ & $16.57$ & $1$\\
NGC7541 & $348.683$ & $4.53436$ & $0.008969$ & $0.032$ & $1.52$ & $2.09$ & $20.08$ & $41.87$ & $1$\\
IC5298 & $349.003$ & $25.5567$ & $0.027422$ & $0.11$ & $0.34$ & $1.95$ & $9.06$ & $11.99$ & $1$\\
NGC7552 & $349.045$ & $-42.5848$ & $0.00536$ & $0.02144$ & $3.76$ & $11.92$ & $77.37$ & $102.92$ & $1$\\
NGC7591 & $349.568$ & $6.58581$ & $0.016531$ & $0.0636$ & $0.28$ & $1.27$ & $7.87$ & $14.87$ & $1$\\
NGC7592 & $349.592$ & $-4.41694$ & $0.024444$ & $0.0972$ & $0.26$ & $0.97$ & $8.05$ & $10.58$ & $1$\\
MRK319 & $349.66$ & $25.2329$ & $0.027012$ & $0.108$ & $0.2211$ & $0.5418$ & $4.266$ & $7.062$ & $2$\\
NGC7673 & $351.921$ & $23.5889$ & $0.01137$ & $0.0422$ & $0.1329$ & $0.5165$ & $4.98$ & $6.893$ & $1,2$\\
NGC7678 & $352.116$ & $22.4212$ & $0.011639$ & $0.0433$ & $0.63$ & $1.16$ & $6.98$ & $14.84$ & $1$\\
MRK534 & $352.194$ & $3.51142$ & $0.01714$ & $0.0677$ & $0.5$ & $1.12$ & $7.4$ & $10.71$ & $1$\\
NGC7679 & $352.194$ & $3.51142$ & $0.017139$ & $0.0662$ & $0.5$ & $1.12$ & $7.58$ & $10.71$ & $1$\\
NGC7714 & $354.059$ & $2.15516$ & $0.00933$ & $0.0386$ & $0.47$ & $2.88$ & $11.16$ & $12.26$ & $1$\\
NGC7771 & $357.854$ & $20.1118$ & $0.01427$ & $0.05711$ & $0.99$ & $2.17$ & $19.67$ & $40.12$ & $1$\\
NGC7793 & $359.458$ & $-32.591$ & $0.00076$ & $0.0031$ & $1.32$ & $1.67$ & $18.14$ & $54.07$ & $1$\\
MRK332 & $359.856$ & $20.7499$ & $0.00802$ & $0.0283$ & $0.3598$ & $0.6212$ & $4.871$ & $9.493$ & $2$\\
\end{longtable}
\end{landscape}
}
\longtabL{2}{
\begin{landscape}
\begin{longtable}{|l|c|c|c|c|c|c|c|c|}
\caption{\label{radio:tab}Radio measurements of the sample. All fluxes in [mJy]. \\References:\\1:~\cite{1991apjs75_1B}, 2:~\cite{2002AJ....124..675C}, 3:~\cite{1983AJ.....88...20C}, 4:~\cite{1998aj115_1693c}, 5:~\cite{1994ApJS...90..179G}, 6:~\cite{1996apjs103_81C}, 7:~\cite{1996ApJS..103..145W}, 8:~\cite{1978ApJS...36...53D}, 9:~\cite{2006AJ....132..546G}, 10:~\cite{1975AJ.....80..771S}, 11:~\cite{2004AJ....127..264B}, 12:~\cite{2007ApJ...654..226R}, 13:~\cite{1995ApJS...97..347G}, 15:~\cite{2004A&A...418....1V}, 16:~\cite{1990pks90C_0000w}, 17:~\cite{1981A&AS...45..367K}, 18:~\cite{1994ApJS...91..111W}, 19:~\cite{1992apjs79_331w}, 20:~\cite{1995apj450_559b}, 21:~\cite{2004ApJ...606..829S}, 22:~\cite{2005A&A...435..521N}, 24:~\cite{1977MNRAS.179..235D}, 25:~\cite{2005ApJ...625..763L}, 26:~\cite{1983ApJS...53..459C}, 27:~\cite{2005ApJS..158....1I}, 28:~\cite{1976ApJ...207..725S}}\\
\hline
\textbf{Name}& $\mathbf{S_{1.40\,GHz}}$ & $\mathbf{S_{2.38\,GHz}}$ & $\mathbf{S_{2.69\,GHz}}$ & $\mathbf{S_{2.70\,GHz}}$ & $\mathbf{S_{4.85\,GHz}}$ & $\mathbf{S_{5.00\,GHz}}$ & $\mathbf{S_{5.01\,GHz}}$ & \textbf{References}\\
\hline\hline
\endfirsthead
\caption{continued.}\\
\hline
\textbf{Name}& $\mathbf{S_{1.40\,GHz}}$ & $\mathbf{S_{2.38\,GHz}}$ & $\mathbf{S_{2.69\,GHz}}$ & $\mathbf{S_{2.70\,GHz}}$ & $\mathbf{S_{4.85\,GHz}}$ & $\mathbf{S_{5.00\,GHz}}$ & $\mathbf{S_{5.01\,GHz}}$ & \textbf{References}\\
\hline\hline
\endhead
\hline
\endfoot
MRK545 & $73.5$ & $47$ & $-$ & $-$ & $33$ & $36$ & $-$ & $1,2,8,14$\\
NGC34 & $67.5$ & $-$ & $-$ & $-$ & $-$ & $-$ & $-$ & $4$\\
MCG-02-01-051 & $43.2$ & $-$ & $-$ & $-$ & $-$ & $-$ & $-$ & $4$\\
NGC174 & $45.7$ & $-$ & $-$ & $-$ & $-$ & $-$ & $-$ & $4$\\
NGC232 & $60.6$ & $-$ & $-$ & $-$ & $56$ & $-$ & $-$ & $4$\\
NGC253 & $6000$ & $-$ & $-$ & $3520$ & $2433$ & $-$ & $2580$ & $5,16,17$\\
IC1623 & $249.2$ & $-$ & $-$ & $-$ & $96$ & $-$ & $-$ & $4,5$\\
NGC520 & $176$ & $110$ & $-$ & $-$ & $87$ & $-$ & $-$ & $1,2,8$\\
NGC632 & $23$ & $15$ & $-$ & $-$ & $-$ & $-$ & $-$ & $2,8$\\
NGC660 & $373$ & $255$ & $-$ & $-$ & $187$ & $-$ & $-$ & $1,4,8$\\
NGC828 & $108$ & $-$ & $-$ & $-$ & $47$ & $-$ & $-$ & $6$\\
NGC891 & $701$ & $-$ & $-$ & $-$ & $342$ & $-$ & $-$ & $2$\\
NGC958 & $71.9$ & $-$ & $-$ & $-$ & $-$ & $-$ & $-$ & $4$\\
NGC1055 & $200.9$ & $129$ & $-$ & $150$ & $63$ & $-$ & $-$ & $1,4,8,16$\\
Maffei2 & $1015$ & $-$ & $-$ & $-$ & $375$ & $-$ & $-$ & $1,19$\\
NGC1068(M77) & $4850$ & $-$ & $-$ & $305$ & $2039$ & $1890$ & $1342.4$ & $2,9,13,14,17$\\
UGC2238 & $72.2$ & $-$ & $-$ & $-$ & $-$ & $-$ & $-$ & $1,2$\\
NGC1097 & $415$ & $-$ & $-$ & $250$ & $126$ & $-$ & $150$ & $6,7,16$\\
NGC1134 & $89.1$ & $57$ & $-$ & $-$ & $32$ & $-$ & $-$ & $1,2,8$\\
NGC1365 & $530$ & $-$ & $-$ & $350$ & $230$ & $180$ & $-$ & $4,7,16$\\
IC342 & $2250$ & $-$ & $-$ & $-$ & $277$ & $-$ & $-$ & $1,6$\\
UGC02982 & $91.3$ & $-$ & $-$ & $-$ & $-$ & $-$ & $-$ & $2$\\
NGC1530 & $80.7$ & $-$ & $-$ & $-$ & $27$ & $-$ & $-$ & $1,6$\\
NGC1569 & $396$ & $-$ & $-$ & $-$ & $198$ & $-$ & $155$ & $1,10,19$\\
MRK617 & $138$ & $-$ & $-$ & $-$ & $63$ & $-$ & $-$ & $4,13$\\
NGC1672 & $450$ & $-$ & $-$ & $210$ & $114$ & $-$ & $100$ & $16,18$\\
MRK1088 & $45.7$ & $31$ & $-$ & $-$ & $-$ & $-$ & $-$ & $2,8$\\
NGC1808 & $497$ & $-$ & $-$ & $350$ & $229$ & $-$ & $220$ & $6,16,18$\\
NGC1797 & $29.1$ & $-$ & $-$ & $-$ & $-$ & $-$ & $-$ & $4$\\
MRK1194 & $42.2$ & $27$ & $-$ & $-$ & $-$ & $-$ & $-$ & $2,8$\\
NGC2146 & $1087$ & $-$ & $-$ & $-$ & $-$ & $-$ & $-$ & $6$\\
NGC2276 & $283$ & $-$ & $-$ & $-$ & $-$ & $-$ & $-$ & $6$\\
NGC2403 & $387$ & $-$ & $-$ & $-$ & $169$ & $-$ & $-$ & $19$\\
NGC2415 & $66.4$ & $53$ & $-$ & $-$ & $41$ & $-$ & $30$ & $1,2,8,10$\\
NGC2782 & $107.5$ & $-$ & $-$ & $-$ & $47$ & $-$ & $60$ & $1,10,20$\\
NGC2785 & $67.6$ & $-$ & $-$ & $-$ & $-$ & $-$ & $-$ & $2$\\
NGC2798 & $82.0$ & $-$ & $-$ & $-$ & $37$ & $-$ & $53$ & $1,2,10$\\
NGC2903 & $444$ & $200$ & $-$ & $-$ & $118$ & $-$ & $-$ & $1,2,8,13$\\
MRK708 & $32.6$ & $21$ & $-$ & $-$ & $-$ & $-$ & $-$ & $2,8$\\
NGC3034(M82) & $7286.8$ & $-$ & $5650$ & $-$ & $3918$ & $-$ & $3912$ & $1,17,20$\\
NGC3079 & $820.7$ & $-$ & $-$ & $-$ & $321$ & $-$ & $-$ & $1,21$\\
NGC3147 & $89.9$ & $-$ & $-$ & $-$ & $44$ & $-$ & $8.1$ & $1,2,23$\\
NGC3256 & $642$ & $-$ & $-$ & $-$ & $319$ & $240$ & $250$ & $5,6,14,16$\\
MRK33 & $24.6$ & $-$ & $-$ & $-$ & $-$ & $-$ & $-$ & $11$\\
NGC3310 & $417$ & $-$ & $-$ & $-$ & $152$ & $-$ & $-$ & $1,19$\\
NGC3367 & $118$ & $71$ & $-$ & $130$ & $36$ & $-$ & $35$ & $1,2,8,10,16$\\
NGC3448 & $51.3$ & $-$ & $-$ & $-$ & $-$ & $-$ & $39$ & $1,10$\\
NGC3504 & $274$ & $-$ & $-$ & $-$ & $117$ & $-$ & $-$ & $1,2,8$\\
NGC3556(M108) & $245$ & $-$ & $-$ & $-$ & $76$ & $-$ & $-$ & $1,19$\\
NGC3627(M66) & $458$ & $209$ & $-$ & $-$ & $141$ & $-$ & $-$ & $1,2,8$\\
NGC3628 & $470.2$ & $313$ & $-$ & $-$ & $276$ & $200$ & $224$ & $1,8,10,14,21$\\
NGC3683 & $127$ & $-$ & $-$ & $-$ & $-$ & $-$ & $-$ & $2$\\
NGC3690 & $658$ & $-$ & $-$ & $-$ & $-$ & $-$ & $362$ & $10,23$\\
MRK188 & $30.7$ & $-$ & $-$ & $-$ & $-$ & $-$ & $25.0$ & $2,10$\\
NGC3893 & $139$ & $-$ & $-$ & $-$ & $39$ & $-$ & $-$ & $1,2$\\
NGC3994 & $70.8$ & $50$ & $-$ & $-$ & $52$ & $-$ & $-$ & $1,2,8$\\
NGC4030 & $147$ & $-$ & $-$ & $90$ & $-$ & $56$ & $-$ & $14,16,19$\\
NGC4041 & $103$ & $-$ & $-$ & $-$ & $48$ & $-$ & $-$ & $1,2$\\
NGC4102 & $273$ & $-$ & $-$ & $-$ & $70$ & $-$ & $105$ & $1,10$\\
MRK1466 & $20.0$ & $15$ & $-$ & $-$ & $-$ & $-$ & $-$ & $2,8$\\
MRK759 & $31.9$ & $16$ & $-$ & $-$ & $13$ & $-$ & $-$ & $2,8,15$\\
NGC4194 & $122$ & $-$ & $-$ & $-$ & $39$ & $-$ & $-$ & $1,19$\\
NGC4214 & $38.3$ & $-$ & $-$ & $-$ & $30$ & $-$ & $-$ & $2$\\
NGC4273 & $78.5$ & $65$ & $-$ & $-$ & $37$ & $-$ & $-$ & $2,8,15$\\
NGC4303(M62) & $444$ & $195$ & $-$ & $-$ & $102$ & $120$ & $-$ & $2,8,14,15$\\
NGC4414 & $227$ & $138$ & $-$ & $-$ & $75$ & $-$ & $-$ & $1,8,19$\\
NGC4418 & $38.5$ & $-$ & $-$ & $-$ & $-$ & $-$ & $-$ & $23$\\
NGC4527 & $187.9$ & $129$ & $-$ & $-$ & $72$ & $-$ & $-$ & $2,8,15$\\
NGC4536 & $204.9$ & $136$ & $-$ & $-$ & $114$ & $110$ & $-$ & $2,8,13$\\
NGC4631 & $1122$ & $340$ & $-$ & $-$ & $438$ & $-$ & $-$ & $1,8,19$\\
NGC4666 & $434$ & $-$ & $-$ & $-$ & $161$ & $-$ & $-$ & $2,13$\\
NGC4793 & $113$ & $72$ & $-$ & $-$ & $46$ & $-$ & $-$ & $1,2,8$\\
NGC4826(M64) & $103$ & $67$ & $-$ & $-$ & $54$ & $-$ & $-$ & $1,2,8$\\
NGC4945 & $6600$ & $-$ & $-$ & $5000$ & $3055$ & $-$ & $2840$ & $16,18$\\
NGC5005 & $194$ & $-$ & $-$ & $-$ & $62$ & $-$ & $-$ & $1,19$\\
NGC5020 & $30.1$ & $22$ & $-$ & $-$ & $-$ & $-$ & $-$ & $2,8$\\
NGC5055(M63) & $349$ & $-$ & $-$ & $-$ & $124$ & $-$ & $-$ & $1,2$\\
ARP193 & $104$ & $-$ & $-$ & $-$ & $53$ & $-$ & $-$ & $1,2$\\
NGC5104 & $39.9$ & $38$ & $-$ & $-$ & $-$ & $-$ & $-$ & $2,8$\\
NGC5135 & $194$ & $-$ & $-$ & $-$ & $107$ & $-$ & $-$ & $6,7$\\
NGC5194(M51) & $1310$ & $-$ & $-$ & $-$ & $436$ & $-$ & $360$ & $1,10,19$\\
NGC5218 & $30.4$ & $-$ & $-$ & $-$ & $-$ & $-$ & $-$ & $2$\\
NGC5236(M83) & $2445$ & $-$ & $-$ & $-$ & $648$ & $-$ & $-$ & $6,7,16$\\
NGC5256 & $159$ & $-$ & $-$ & $-$ & $47$ & $-$ & $-$ & $1,19$\\
NGC5257 & $48.7$ & $48$ & $-$ & $-$ & $-$ & $-$ & $-$ & $2,8$\\
NGC5253 & $83.8$ & $-$ & $-$ & $-$ & $90$ & $75$ & $-$ & $6,7,14$\\
UGC8739 & $93.6$ & $-$ & $-$ & $-$ & $37$ & $-$ & $-$ & $1,2$\\
MRK1365 & $23.0$ & $14$ & $-$ & $-$ & $-$ & $-$ & $-$ & $2,8$\\
NGC5430 & $65.9$ & $-$ & $-$ & $-$ & $29$ & $-$ & $40$ & $1,2,10$\\
NGC5427 & $63$ & $-$ & $-$ & $-$ & $-$ & $-$ & $-$ & $26$\\
NGC5678 & $110$ & $-$ & $-$ & $-$ & $68$ & $-$ & $-$ & $2$\\
NGC5676 & $116$ & $-$ & $-$ & $-$ & $38$ & $-$ & $33$ & $2$\\
NGC5713 & $222$ & $-$ & $-$ & $-$ & $93$ & $73$ & $-$ & $1,2,14$\\
NGC5775 & $221$ & $138$ & $-$ & $-$ & $67$ & $-$ & $-$ & $1,8,19$\\
NGC5900 & $60.4$ & $-$ & $-$ & $-$ & $-$ & $17$ & $-$ & $2,10$\\
NGC5936 & $139$ & $81$ & $-$ & $-$ & $48$ & $-$ & $59$ & $1,2,8,10$\\
ARP220 & $324$ & $-$ & $260$ & $-$ & $208$ & $-$ & $-$ & $1,2,3$\\
NGC5962 & $82.3$ & $56$ & $-$ & $-$ & $36$ & $-$ & $-$ & $1,2,8$\\
NGC5990 & $63.9$ & $39$ & $-$ & $-$ & $-$ & $-$ & $-$ & $2,8$\\
NGC6181 & $95.6$ & $60$ & $-$ & $-$ & $56$ & $-$ & $-$ & $1,2,8$\\
NGC6217 & $79.9$ & $-$ & $-$ & $-$ & $-$ & $-$ & $-$ & $2$\\
NGC6240 & $653$ & $-$ & $-$ & $-$ & $179$ & $-$ & $170$ & $13,16,19$\\
NGC6286 & $157$ & $-$ & $-$ & $-$ & $53$ & $-$ & $-$ & $1,2$\\
IRAS18293-3413 & $223$ & $-$ & $-$ & $-$ & $144$ & $-$ & $-$ & $6,7$\\
NGC6701 & $92.2$ & $-$ & $-$ & $-$ & $22$ & $-$ & $-$ & $1,6$\\
NGC6764 & $115.0$ & $-$ & $-$ & $-$ & $34$ & $-$ & $47$ & $1,2,10$\\
NGC6946 & $1395$ & $-$ & $-$ & $-$ & $531$ & $-$ & $-$ & $1,19$\\
NGC7130 & $183$ & $-$ & $-$ & $-$ & $-$ & $-$ & $-$ & $6$\\
IC5179 & $165$ & $-$ & $-$ & $-$ & $79$ & $-$ & $-$ & $6,7$\\
NGC7331 & $373$ & $187$ & $-$ & $-$ & $80$ & $-$ & $-$ & $1,2,8$\\
NGC7469 & $255$ & $132$ & $-$ & $-$ & $95$ & $-$ & $-$ & $8,13,19$\\
NGC7479 & $107$ & $58$ & $-$ & $-$ & $41$ & $-$ & $-$ & $1,2,8$\\
NGC7496 & $36.3$ & $-$ & $-$ & $-$ & $-$ & $-$ & $-$ & $6$\\
NGC7541 & $162$ & $101$ & $-$ & $-$ & $56$ & $-$ & $-$ & $1,2,8$\\
IC5298 & $24.2$ & $-$ & $-$ & $-$ & $-$ & $-$ & $-$ & $23$\\
NGC7552 & $276$ & $-$ & $-$ & $-$ & $139$ & $-$ & $-$ & $6,18$\\
NGC7591 & $52.1$ & $39$ & $-$ & $-$ & $-$ & $-$ & $-$ & $2,8$\\
NGC7592 & $75$ & $-$ & $-$ & $-$ & $-$ & $-$ & $-$ & $27$\\
MRK319 & $31.6$ & $21$ & $-$ & $-$ & $-$ & $-$ & $-$ & $2,8$\\
NGC7673 & $43.4$ & $30$ & $-$ & $-$ & $-$ & $-$ & $-$ & $2,8$\\
NGC7678 & $49.5$ & $36$ & $-$ & $-$ & $-$ & $-$ & $-$ & $2,8$\\
MRK534 & $55.8$ & $33$ & $-$ & $-$ & $45$ & $-$ & $-$ & $2,8,13$\\
NGC7679 & $55.8$ & $33$ & $-$ & $-$ & $45$ & $-$ & $-$ & $2,8,13$\\
NGC7714 & $65.8$ & $-$ & $-$ & $-$ & $39$ & $-$ & $-$ & $1,2$\\
NGC7771 & $229$ & $97$ & $-$ & $-$ & $57$ & $-$ & $-$ & $1,8,19$\\
NGC7793 & $103$ & $-$ & $-$ & $-$ & $-$ & $-$ & $-$ & $6$\\
MRK332 & $36.5$ & $27$ & $-$ & $-$ & $-$ & $-$ & $-$ & $2,8$\\
\end{longtable}
\end{landscape}
}

\longtabL{3}{
\begin{landscape}
\begin{longtable}{|l|c|c|c|c|c|c|c|c|}
\caption{\label{xray:tab}X-Ray measurements of the sample. All fluxes in [nJy]. \\References:\\1:~\cite{1996MNRAS.278.1049R}, 2:~\cite{1992ApJS...80..531F}, 3:~\cite{2000yCat.9031....0W}, 4:~\cite{2005A&A...435..799T}, 5:~\cite{1994A&A...281..355B}, 6:~\cite{2005MNRAS.358.1423O}, 7:~\cite{2006A&A...448...43T}, 8:~\cite{2005ApJ...633..664T}, 9:~\cite{2007ApJ...657..167S}, 10:~\cite{2005A&A...444..119G}}\\
\hline
\textbf{Name}& \textbf{EO IPC(ROSAT)} & \textbf{EINSTEIN} & \textbf{Chandra} & \textbf{ROSAT} & \textbf{ROSAT} & \textbf{Chandra} & \textbf{XMM} & \textbf{References}\\
~& \textbf{0.1-4\,keV} &  \textbf{0.2-4.0\,keV} & \textbf{0.1-2.4\,keV} & \textbf{0.2-2.0\,keV} & \textbf{0.1-2.4\,keV} & \textbf{0.3-8\,keV} & \textbf{0.3-2\,keV} & ~ \\\hline\hline
\endfirsthead
\caption{continued.}\\
\hline
\textbf{Name}& \textbf{EO IPC(ROSAT)} & \textbf{EINSTEIN} & \textbf{Chandra} & \textbf{ROSAT} & \textbf{ROSAT} & \textbf{Chandra} & \textbf{XMM} & \textbf{References}\\
~& \textbf{0.1-4\,keV} &  \textbf{0.2-4.0\,keV} & \textbf{0.1-2.4\,keV} & \textbf{0.2-2.0\,keV} & \textbf{0.1-2.4\,keV} & \textbf{0.3-8\,keV} & \textbf{0.3-2\,keV} & ~ \\\hline\hline
\endhead
\hline
\endfoot
NGC34 & $-$ & $-$ & $-$ & $-$ & $-$ & $-$ & $23$ & $9$\\
NGC520 & $-$ & $-$ & $-$ & $33.4$ & $-$ & $-$ & $-$ & $3$\\
NGC660 & $-$ & $-$ & $-$ & $45.1$ & $-$ & $-$ & $-$ & $3$\\
NGC891 & $-$ & $-$ & $-$ & $-$ & $-$ & $-$ & $69.4$ & $7$\\
NGC1068(M77) & $-$ & $3940$ & $-$ & $-$ & $10900$ & $-$ & $-$ & $2,5$\\
NGC1097 & $-$ & $578$ & $-$ & $-$ & $-$ & $-$ & $-$ & $2$\\
NGC1365 & $-$ & $326$ & $-$ & $-$ & $-$ & $-$ & $-$ & $2$\\
IC342 & $-$ & $939$ & $-$ & $-$ & $-$ & $-$ & $-$ & $2$\\
NGC1569 & $-$ & $368$ & $-$ & $-$ & $-$ & $-$ & $-$ & $2$\\
MRK617 & $-$ & $124$ & $-$ & $-$ & $-$ & $-$ & $-$ & $2$\\
MRK1088 & $-$ & $-$ & $-$ & $42.3$ & $-$ & $-$ & $-$ & $8$\\
NGC1808 & $-$ & $-$ & $-$ & $-$ & $446$ & $-$ & $-$ & $5$\\
NGC2403 & $-$ & $364$ & $-$ & $-$ & $-$ & $-$ & $-$ & $2$\\
NGC2415 & $-$ & $-$ & $-$ & $63.8$ & $-$ & $-$ & $-$ & $3$\\
NGC2903 & $-$ & $273$ & $-$ & $46.4$ & $-$ & $-$ & $-$ & $2,3$\\
NGC3034(M82) & $-$ & $4490$ & $-$ & $-$ & $-$ & $-$ & $-$ & $2$\\
NGC3079 & $-$ & $113$ & $-$ & $-$ & $-$ & $-$ & $-$ & $2$\\
NGC3256 & $-$ & $-$ & $-$ & $-$ & $1060$ & $-$ & $-$ & $5$\\
NGC3310 & $-$ & $208$ & $-$ & $-$ & $-$ & $-$ & $-$ & $2$\\
NGC3367 & $102$ & $-$ & $-$ & $-$ & $-$ & $-$ & $-$ & $1$\\
NGC3448 & $-$ & $72.6$ & $-$ & $-$ & $-$ & $-$ & $-$ & $2$\\
NGC3504 & $-$ & $76.8$ & $-$ & $-$ & $-$ & $-$ & $-$ & $2$\\
NGC3627(M66) & $-$ & $-$ & $-$ & $-$ & $603$ & $-$ & $-$ & $5$\\
NGC3690 & $-$ & $91.5$ & $-$ & $-$ & $-$ & $-$ & $-$ & $2$\\
NGC4102 & $-$ & $-$ & $-$ & $-$ & $-$ & $-$ & $45.1$ & $9$\\
NGC4214 & $-$ & $-$ & $-$ & $-$ & $-$ & $75.6$ & $-$ & $6$\\
NGC4273 & $-$ & $-$ & $-$ & $67.1$ & $-$ & $-$ & $-$ & $3$\\
NGC4303(M62) & $-$ & $180$ & $-$ & $-$ & $233$ & $-$ & $-$ & $2,5$\\
NGC4536 & $-$ & $112$ & $-$ & $-$ & $-$ & $-$ & $-$ & $2$\\
NGC4631 & $-$ & $263$ & $-$ & $-$ & $-$ & $-$ & $-$ & $2$\\
NGC4826(M64) & $-$ & $150$ & $-$ & $53.3$ & $-$ & $-$ & $-$ & $3$\\
NGC4945 & $-$ & $-$ & $-$ & $-$ & $794$ & $-$ & $-$ & $5$\\
NGC5135 & $-$ & $63.9$ & $-$ & $176$ & $-$ & $-$ & $-$ & $3$\\
NGC5236(M83) & $-$ & $972$ & $-$ & $-$ & $954$ & $-$ & $-$ & $2,5$\\
NGC5256 & $-$ & $-$ & $-$ & $-$ & $-$ & $-$ & $9.93$ & $10$\\
NGC5253 & $-$ & $41.4$ & $-$ & $38.5$ & $-$ & $35.8$ & $-$ & $2,3,6$\\
NGC5775 & $-$ & $-$ & $-$ & $-$ & $-$ & $-$ & $15.8$ & $7$\\
ARP220 & $45.69$ & $-$ & $-$ & $-$ & $-$ & $-$ & $-$ & $1$\\
NGC6240 & $-$ & $-$ & $-$ & $-$ & $914$ & $-$ & $-$ & $5$\\
NGC6946 & $-$ & $611$ & $-$ & $-$ & $-$ & $-$ & $-$ & $2$\\
NGC7331 & $-$ & $16.8$ & $-$ & $76.7$ & $-$ & $-$ & $-$ & $2,3$\\
NGC7469 & $-$ & $12000$ & $-$ & $-$ & $-$ & $-$ & $-$ & $2$\\
NGC7552 & $-$ & $167$ & $-$ & $-$ & $-$ & $-$ & $-$ & $2$\\
MRK534 & $-$ & $204$ & $-$ & $-$ & $-$ & $-$ & $-$ & $3$\\
NGC7679 & $-$ & $204$ & $-$ & $-$ & $-$ & $-$ & $-$ & $2$\\
NGC7714 & $-$ & $53.8$ & $-$ & $-$ & $-$ & $-$ & $-$ & $2$\\
NGC7771 & $-$ & $138$ & $-$ & $-$ & $-$ & $-$ & $-$ & $2$\\
NGC7793 & $-$ & $130$ & $-$ & $-$ & $-$ & $-$ & $-$ & $2$\\
\end{longtable}
\end{landscape}
}
\end{appendix}

\end{document}